\documentclass[prd,twocolumn,showpacs,preprintnumbers,amsmath,nofootinbib,amssymb,superscriptaddress]{revtex4}

\usepackage{bbm,lmodern,amsmath,graphicx,epsfig,amssymb,amsfonts,dsfont,mathtools}
\usepackage{amsmath,amssymb,mathrsfs}
\usepackage{rotating}
\usepackage{bigstrut}
\usepackage{morefloats}
\usepackage{multirow}
\usepackage{longtable}
\usepackage{dcolumn}
\usepackage{placeins}
\usepackage{ulem}

\usepackage{hyperref}
\hypersetup{
	colorlinks	=true,
	urlcolor	=blue,
	linkcolor	=blue,
	citecolor	=blue,
	pdftitle	={Eta photoproduction in a combined analysis of pion- and photon-induced reactions},
	pdfauthor	={D.~R\"onchen, M.~D\"oring, H.~Haberzettl, J.~Haidenbauer, U.-G.~Mei\ss ner, K. Nakayama},
	pdfsubject	={Eta photoproduction in a combined analysis of pion- and photon-induced reactions}
	}

\allowdisplaybreaks[1]

\newcommand{\be}{\begin{equation}}
\newcommand{\ee}{\end{equation}}
\newcommand{\ba}{\begin{eqnarray}}
\newcommand{\ea}{\end{eqnarray}}
\newcommand{\non}{\nonumber \\}
\newcommand{\po}{{\rm P}}
\newcommand{\npo}{{\rm NP}}
\begin{document}

\title{Eta photoproduction in a combined analysis of pion- and photon-induced reactions}

\author{D.~R\"onchen}
\email{roenchen@hiskp.uni-bonn.de}
\affiliation{Helmholtz-Institut f\"ur Strahlen- und Kernphysik (Theorie) and Bethe Center for Theoretical
Physics,  Universit\"at Bonn, 53115 Bonn, Germany}

\author{M.~D\"oring}
\email{doring@gwu.edu}
\affiliation{Institute for Nuclear Studies and Department of Physics, The George Washington University,
Washington, DC 20052, USA}

%\author{F.~Huang}
%\affiliation{School of Physics, University of Chinese Academy of Sciences, Huairou District, Beijing 101408, China}
%\affiliation{Department of Physics and Astronomy, University of Georgia, Athens, Georgia 30602, USA}

\author{H.~Haberzettl}
\affiliation{Institute for Nuclear Studies and Department of Physics, The George Washington University,
Washington, DC 20052, USA}

\author{J.~Haidenbauer}
\affiliation{Institut f\"ur Kernphysik and J\"ulich Center for Hadron Physics, Forschungszentrum J\"ulich, 
52425 J\"ulich, Germany}
\affiliation{Institute for Advanced Simulation, Forschungszentrum J\"ulich, 52425 J\"ulich, Germany}

%\author{C.~Hanhart}
%\affiliation{Institut f\"ur Kernphysik and J\"ulich Center for Hadron Physics, Forschungszentrum J\"ulich, 
%52425 J\"ulich, Germany}
%\affiliation{Institute for Advanced Simulation, Forschungszentrum J\"ulich, 52425 J\"ulich, Germany}

%\author{S.~Krewald} 
%\affiliation{Institut f\"ur Kernphysik and J\"ulich Center for Hadron Physics, Forschungszentrum J\"ulich, 
%52425 J\"ulich, Germany}
%\affiliation{Institute for Advanced Simulation, Forschungszentrum J\"ulich, 52425 J\"ulich, Germany}

\author{U.-G.~Mei\ss ner}
\affiliation{Helmholtz-Institut f\"ur Strahlen- und Kernphysik (Theorie) and Bethe Center for Theoretical
Physics,  Universit\"at Bonn, 53115 Bonn, Germany}
\affiliation{Institut f\"ur Kernphysik and J\"ulich Center for Hadron Physics, Forschungszentrum J\"ulich, 
52425 J\"ulich, Germany}
\affiliation{Institute for Advanced Simulation, Forschungszentrum J\"ulich, 52425 J\"ulich, Germany}

\author{K. Nakayama}
\affiliation{Institut f\"ur Kernphysik and J\"ulich Center for Hadron Physics, Forschungszentrum J\"ulich, 
52425 J\"ulich, Germany}
\affiliation{Department of Physics and Astronomy, University of Georgia, Athens, Georgia 30602, USA}

\begin{abstract}
The $\eta N$ final state is
isospin-selective and thus provides access to the spectrum of excited
nucleons without being affected by excited $\Delta$ states. To this end, the
world database on eta photoproduction off the proton up to a center-of-mass energy of
$E\sim 2.3$~GeV is analyzed, including data on differential cross sections,
and single and double polarization observables. The
resonance spectrum and its properties are determined in a combined analysis
of eta and pion photoproduction off the proton together with the reactions
$\pi N\to \pi N$, $\eta N$, $K\Lambda$ and $K\Sigma$. For
the analysis, the so-called J\"ulich coupled-channel framework is used,
incorporating unitarity, analyticity, and effective three-body channels. Parameters tied
to photoproduction and hadronic interactions are varied simultaneously.
The influence of recent MAMI $T$ and $F$ asymmetry data on the eta photoproduction amplitude is discussed in detail. 
\end{abstract}

\pacs{
{11.80.Gw}, % Multichannel scattering
{13.60.Le}, % Photoproduction of mesons
{13.75.Gx}. % Pion-baryon interactions
}

\maketitle

%%%%%%%%%%%%%%%%%%%%%%%%%%%%%%%%%%%%%%%%%%%%%%%%%%%%%%%%%%%%%%%%%%%%%%%%%%%%%%%%%%%%%%%%%%%%%%%%%%%%%%%%%%%%%

\section{Introduction}

The determination of the spectrum of excited baryons from experimental data is necessary to understand Quantum Chromodynamics at low and medium energies. In this non-perturbative regime, quark models~\cite{Capstick:1993kb,Ronniger:2011td} and lattice calculations~\cite{Edwards:2012fx,Lang:2012db,Engel:2013ig} predict more exited states than found so far in partial-wave analyses of experimental data. This dilemma is known as ``missing resonance problem"~\cite{Koniuk:1979vw}. 
In the past, the dominant source of information on resonance properties was provided by elastic $\pi N$ scattering~\cite{Cutkosky:1979fy,Hoehler1,Arndt:2006bf}. 
The analysis of inelastic reactions is, however, essential~\cite{Ronchen:2012eg,Burkert:2014wea} when aiming at a reliable extraction of the entire spectrum and, consequently, at an identification of missing states that might couple predominantly to channels other than $\pi N$. 

Among the inelastic channels accessible in $\pi N$ scattering,  the $\eta N$ channel plays a crucial role. 
It couples exclusively to states with isospin $I=1/2$ allowing for the extraction of $N^*$ states unaffected by contributions from $\Delta^*$ states. Moreover, the $\eta N$ channel opens at relatively low energies, in a region which is populated by numerous nucleon resonances. For example, some states like the four-star $N(1535) 1/2^-$ resonance are known to have a large $\eta N$ branching ratio. Less well-established resonances like the $N(1710) 1/2^+$, whose parameters are only weakly constrained from elastic $\pi N$ scattering~\cite{Arndt:2006bf,Burkert:2014wea,Ceci:2011ae,Ceci:2006ra}, may show a noticeable signal in their $\eta N$ decay~\cite{Ronchen:2012eg, Burkert:2014wea}. A narrow structure discovered in eta photoproduction on the neutron~\cite{Kuznetsov:2006kt, Miyahara:2007zz, Werthmuller:2013rba} at around $E=1.68$~GeV could also appear in eta production on the proton target.

In this respect it is unfortunate that the database for the reaction $\pi N\to \eta N$ is problematic over the whole energy  range. Its coverage in scattering angles and energies is too limited to perform a well-founded resonance analysis. Furthermore, for energies around 100 MeV from the threshold and beyond, only some experiments were performed and many of those are known to suffer from systematic uncertainties~\cite{Clajus:1992, Arndt:2006bf, Ronchen:2012eg}. 

An alternative experimental window has opened in recent years with high-precision measurements of cross sections and polarization observables in eta photoproduction at photon-beam facilities like ELSA, JLab or MAMI, see Refs.~\cite{Krusche:2014ava,Aznauryan:2011qj,Klempt:2009pi} for reviews. 
For example, the first measurements from the JLab FROST target (polarization $E$ for $\gamma p\to\pi^+n$) have appeared only recently~\cite{Strauch:2015zob}, and many more polarization data are expected.
The database for eta photoproduction is not yet as large as the one for pion photoproduction, but it is rapidly growing. In addition, the data already available are of much higher quality than those for the pion-induced reaction $\pi N\to \eta N$. Recently, the first data for the beam-target asymmetry $F$ in $\gamma p\to \eta p$ was presented by the A2 collaboration at MAMI, together with a measurement of the target asymmetry $T$~\cite{Akondi:2014ttg}.

A key question is whether eta photoproduction data can be used to access the poorly known $N^*$ branching ratios to the $\eta N$ channel. The combined analysis of elastic $\pi N$ scattering and pion photoproduction determines, at least in principle, the helicity couplings and $\pi N$ branching ratios. With known helicity couplings, then indeed the data on eta photoproduction allow one to pin down the resonance $\eta N$ branching ratios, without having to resort to the problematic $\pi N\to\eta N$ data~\cite{Clajus:1992, Arndt:2006bf, Ronchen:2012eg}. This argumentation is, however, limited by the fact that even for pion photoproduction the database is not yet fully complete. The extension of the analysis to eta photoproduction will then improve the knowledge of the $\eta N$ branching ratios, but, due to incomplete databases, will also lead to changes both in the helicity couplings and branching ratios, as will be seen. 

Given the discussed mismatch in data quality, it is not recommendable to only fit model parameters tied to photoproduction and leave the hadronic interaction unchanged: The hadronic amplitude, poorly fixed from the $\pi N\to\eta N$ data, appears as a sub-process in photoproduction. Thus, to avoid any bias, an unconstrained fit is needed. In this way, the higher statistical weight of the eta photoproduction data even provides a 
better constraint on the hadronic $\pi N\to\eta N$ amplitude.

The photoproduction of $\eta$ mesons is also a prime candidate for a ``complete experiment''.
From a mathematical point of view, a complete experiment~\cite{Barker:1975bp} consists of a set of eight carefully chosen observables, which resolve all discrete ambiguities up to an overall phase~\cite{Chiang:1996em, 	Keaton:1996pe}. For example, a complete set including $F$ and $T$ is given by $\lbrace \sigma, \Sigma, T, P, E, F, C_x, O_x\rbrace$~\cite{Chiang:1996em}. For experiments with realistic uncertainties, however, eight observables are not sufficient~\cite{Sandorfi:2010uv,Vrancx:2013pza,Nys:2015kqa}. Less than eight observables are required in a truncated partial-wave analysis~\cite{Omelaenko, Wunderlich:2014xya}.

By contrast, for a complete experiment on the reaction $\pi^-p\to\eta n$ only four observables are needed~\cite{Jackson:2013bba, Jackson:2015dva}.
Given the data situation for that reaction, a re-measurement would greatly advance our understanding of the $\eta N$ final state. Physics opportunities with a pion beam are discussed in Ref.~\cite{hadwp}. In any case, the current database does not contain a complete set of observables, neither for photon- nor pion-induced eta production. One then has to resort to other approaches that often combine data from different initial and final states.

Over the years, a variety of theoretical approaches has been applied to analyze the pion- and photon-induced production of $\eta$ mesons. For example, photoproduction of $\eta$ mesons in the resonance region not too far from threshold was studied in the framework of unitarized chiral perturbation theory in Refs.~\cite{Kaiser:1996js,Inoue:2001ip,Borasoy:2002mt,Doring:2009qr,Mai:2012wy}.
Considering a broader energy range up to and beyond 2~GeV, $K$-matrix~\cite{Arndt:2005dg, Anisovich:2012ct,Shklyar:2012js,Shrestha:2012ep,Batinic:2010zz, Cutkosky:1979fy,Gridnev:1999sz}  and unitary isobar~\cite{Chiang:2001as} models are practical tools to perform an analysis of large amounts of data. Sometimes, the real part of the self-energies is neglected and only on-shell intermediate states are maintained, which reduces the complexity of the calculations.
For the purpose of a combined analysis of different reactions over a wide energy range, so-called dynamical coupled-channel (DCC) models provide a particularly suited framework. Theoretical constraints of the $S$-matrix, like two- and three-body unitarity, analyticity, left-hand cuts and complex branch points, are manifestly implemented or at least approximated. This enables the reliable determination of the resonance spectrum in terms of pole positions, residues, and helicity couplings in the complex energy plane. The production of $\eta$ mesons in DCC approaches was studied, e.g., in Refs.~\cite{Kamano:2013iva, Chen:2007cy}. 

Here, we extend the J\"ulich model, a DCC approach pursued over many years~\cite{Doring:2009bi, Doring:2009yv, Doring:2010ap,Ronchen:2012eg,Ronchen:2014cna} starting with Ref.~\cite{Schutz:1994ue}, to perform a simultaneous analysis of the pion-induced reactions $\pi N\to \pi N$, $\pi^-p\to\eta n$, $K^0\Lambda$, $K^+\Sigma^-$, $K^0\Sigma^0$, $\pi^+p\to K^+\Sigma^+$, and the photon-induced reactions $\gamma p\to \pi^0p$, $\pi^+n$, and $\eta p$.
We allow the hadronic amplitudes themselves to
vary, in addition to the parameters tied to photoproduction. As discussed, this is necessary because the quality of the data in $\pi^-p\to\eta n$ is much inferior to the data in eta photoproduction.
In a simultaneous fit to all pion- and photon-induced data, we observe that, indeed, the eta photoproduction data have a strong
influence. This influence reaches beyond the electromagnetic resonance properties and affects also resonance pole positions
and hadronic branching ratios.

The paper is organized as follows: In Sec.~\ref{formalism}, a short overview of the applied formalism is given. For a more detailed introduction of the semi-phenomenological approach to meson photoproduction we refer the reader to Ref.~\cite{Ronchen:2014cna}. In Ref.~\cite{Ronchen:2012eg}, an extensive description of the hadronic J\"ulich DCC framework is provided. In Sec.~\ref{sec:data_fit} we describe the data analysis and in Sec.~\ref{sec:fit} the fit results are shown. The extracted resonance parameters are presented and discussed in Sec.~\ref{resonances}. Technical details about the renormalization of the nucleon mass are summarized in an appendix.

\section{Formalism}
\label{formalism}
In the approach (referred to as ``J\"ulich model''), the hadronic scattering potential is iterated in a Lippmann-Schwinger equation formulated in time-ordered perturbation theory (TOPT) and two-body unitarity is, thus, automatically fulfilled. The three-body $\pi\pi N$ states are parameterized through the channels $\rho N$, $\sigma N$ and $\pi \Delta$. These effective three-body channels are included dynamically, i.e., the $\pi\pi$ and $\pi N$ subsystems match the corresponding phase shifts~\cite{Doring:2009yv}. The analytic structure of the amplitude is given through real and complex branch points~\cite{Ceci:2011ae} and the real, dispersive contributions of the intermediate states. Moreover, $t$- and $u$-channel exchanges of known mesons and baryons constitute the non-resonant part of the amplitude and serve as ``background". 
While the $u$-channel diagrams approximate the left-hand cuts, $t$-channel meson exchanges are essential to achieve three-body unitarity~\cite{Aaron:1969my}. Note that the latter is, at the moment, only approximately satisfied in the J\"ulich model.
By means of this explicit treatment of the background, strong correlations between the different partial waves and a non-trivial energy and angular dependence of the observables are generated. Although $t$- and $u$-channel processes are necessary for analytic structure and unitarity, they do not fully determine the
amplitude. Bare resonance states are included as $s$-channel processes. In contrast to previous versions of the approach, here we also allow for additional contact interactions. Such
interactions do not spoil the analytic properties ensured by $s$-, $t$- and $u$-channel interactions. They absorb physics beyond
the explicit processes and, thus, increase the model-independence of the approach at the cost of a few more parameters. Practically, the changes in the amplitudes induced by the contact terms are comparatively small and the so-called background is still dominated by the $t$- and $u$-channel exchanges. Details are given in the following section. Note that contact terms are also included in Ref.~\cite{Jackson:2015dva}.

In Ref.~\cite{Doring:2010ap} the approach was extended to the strangeness sector incorporating the $K^+\Sigma^+$ final state in the analysis.
In the J\"ulich2012 model of Ref.~\cite{Ronchen:2012eg}, the spectrum of nucleon and $\Delta$ resonances was extracted from a simultaneous analysis of the reactions $\pi N\to\pi N$, $\eta N$, $K\Lambda$ and $K\Sigma$.   
The extension of the J\"ulich approach to pion photoproduction in a field-theoretical formulation, that respects the generalized off-shell Ward-Takahashi identity, was achieved in Ref.~\cite{Huang:2011as}. In Ref.~\cite{Ronchen:2014cna}, by contrast, the photon interaction is approximated in a phenomenological framework and the J\"ulich2012 analysis serves as final-state interaction. The flexible 
formulation of Ref.~\cite{Ronchen:2014cna}, used to study the world data on pion photoproduction on the proton, proved to be capable of analyzing large amounts of data while at the same time maintaining the analytic properties of the J\"ulich approach. This framework, with the addition of contact terms, will be applied in the present study.

%%%%%%%%%%%%%%%%%%%%%%%%%%%%%%%%%%%%%%%%%%%%%%%%%%%%%%%%%%%%%%%%%

\subsection{Pion-induced reactions}
\label{formalism_pion}

The pion-induced reactions are treated within the J\"ulich dynamical coupled-channel formalism~\cite{Ronchen:2012eg}. The $T$-matrix which describes the scattering process of a baryon and a meson can be formulated in the partial-wave basis and reads
\begin{multline}
T_{\mu\nu}(q,p',E)=V_{\mu\nu}(q,p',E) \\
+\sum_{\kappa}\int\limits_0^\infty dp\,
 p^2\,V_{\mu\kappa}(q,p,E)G^{}_\kappa(p,E)\,T_{\kappa\nu}(p,p',E) \ .
\label{scattering}
\end{multline}
In Eq.~(\ref{scattering}) and in the following, $E$ always means the scattering energy in the center-of-mass (c.m.) frame and $q\equiv|\vec q\,|$ ($p'\equiv |\vec p\,'|$) is the
 modulus of the outgoing (incoming)  three-momentum that can be on- or off-shell. The channel indices $\nu$, $\mu$ and $\kappa$ represent the incoming, 
 outgoing and intermediate meson-baryon pairs, respectively.

The propagators $G_\kappa$ for channels with stable particles, i.e. $\kappa=\pi N$, $\eta N$, $K\Lambda$, or $K\Sigma$,  are given by 
\begin{equation}
G_\kappa(p,E)=\frac{1}{E-E_a(p)-E_b(p)+i\epsilon}\;,
\end{equation}
with $E_a=\sqrt{m_a^2+p^2}$ and  $E_b=\sqrt{m_b^2+p^2}$ being the on-mass-shell energies of the
intermediate particles $a$ and $b$ in channel $\kappa$ with respective masses $m_a$ and $m_b$. In case of the channels with unstable particles $\rho N$, $\sigma N$ and $\pi\Delta$ that parameterize the $\pi\pi N$ channels, the propagators are more
complex; for details see Ref.~\cite{Doring:2009yv}.

The scattering potentials $V_{\mu\nu}$ can be decomposed into a pole and a non-pole part
\be
V_{\mu\nu}=V^\npo_{\mu\nu}+V^\po_{\mu\nu}\equiv V^\npo_{\mu\nu}+\sum_{i=0}^{n} 
\frac{\gamma^a_{\mu;i}\,\gamma^c_{\nu;i}}{E-m_i^b} \;. 
\label{vblubb}
\ee
The quantity $V^\npo$ denotes the sum of all $t$- and $u$-channel exchange diagrams, while $V^\po$ comprises the $s$-channel resonance graphs.
The functions $\gamma^c_{\mu;i}$
($\gamma^a_{\nu;i}$) correspond to the bare \underline{c}reation (\underline{a}nnihilation) vertices of a resonance $i$ with bare mass $m_i^b$ and are constructed from an effective Lagrangian which can be found in Table~8 of Ref.~\cite{Doring:2010ap}. Explicit expressions of $\gamma^c_{\mu;i}$ and $\gamma^a_{\nu;i}$ are also given in Ref.~\cite{Doring:2010ap}, cf. also Appendix~A of Ref.~\cite{Ronchen:2012eg}. The exchange potentials that constitute $V^\npo$, also derived from effective Lagrangians, are compiled in Appendix~B of Ref.~\cite{Ronchen:2012eg}.
The decomposition of Eq.~(\ref{vblubb}) is slightly modified in the present approach. We implement here contact terms that do not introduce any singularities and that are used to absorb physics not explicitly contained in the parameterization through
$s$-, $t$- and $u$-channel processes. The contact terms are introduced in a separable form and separately for every partial wave,
\be
V_{\mu\nu}^{{\rm CT}}=\frac{1}{m_N}\gamma^{{\rm CT};a}_{\mu}\gamma^{{\rm CT};c}_\nu\;,
\label{ctct}
\ee
where the $\gamma^{{\rm CT};c}_\mu$ ($\gamma^{{\rm CT};a}_\mu$) have the same functional form as the resonance vertices $\gamma^c_{\mu;i}$ ($\gamma^a_{\mu;i}$) in Eq.~(\ref{vblubb}). The associated couplings in the $\gamma^{{\rm CT}}$ are now new free parameters to be adapted in the fit to the data. 

Formally, the numerator structure of Eq.~(\ref{ctct}) is the same as the one of $s$-channel pole terms, to ensure the correct threshold behavior. Also,
the contact terms carry channel indices. Formally, we can
treat the contact terms as bare $s$-channel processes and absorb the contributions in the
definition of $V^{{\rm P}}$:
\be
V_{\mu\nu}^{{\rm P}}\to \sum_{i=0}^{n} \frac{\gamma^a_{\mu;i}\,\gamma^c_{\nu;i}}{E-m_i^b}+ V_{\mu\nu}^{{\rm CT}}\;.
\ee
For a compact notation, we will no longer distinguish between bare resonance vertices
$\gamma$ and structures of the contact term $\gamma^{{\rm CT}}$ in the following. Then, the index $i$ can refer to a bare $s$-channel resonance vertex $\gamma$ or one of the terms in the numerator of Eq.~(\ref{ctct}).

Similar to the potential $V_{\mu\nu}$, the scattering matrix $T_{\mu\nu}$ can also be written as the sum of a pole and a non-pole part,
 \begin{equation}
T_{\mu\nu}=T_{\mu\nu}^\po+T_{\mu\nu}^\npo 
\label{deco1} 
\end{equation} 
with the unitary $T_{\mu\nu}^\npo$ defined through
 \begin{equation}
T^\npo_{\mu\nu}=V_{\mu\nu}^\npo+\sum_\kappa V_{\mu\kappa}^\npo G_{\kappa} T_{\kappa\nu}^\npo \ .
\label{deco2} 
\end{equation} 
Here and in the following, we do not display function arguments and integration symbols in favor of a more compact notation. 
Besides the resonance poles arising from $s$-channel diagrams in $T^\po$, a dynamical generation of poles in $T^\npo$ is also possible, as explained in Refs.~\cite{Doring:2009bi, Ronchen:2012eg}. 

%$n$ is the number of bare $s$-channel states in a given partial wave

Note that the pole part $T^\po$ can be evaluated from the non-pole part $T^\npo$. To this purpose one defines the dressed creation (annihilation) vertex  $\Gamma_{\mu;i}^c$ ($\Gamma_{\mu;i}^a$) via
\begin{eqnarray}
\Gamma_{\mu;i}^c	&=&\gamma^c_{\mu;i}+\sum_\nu  \gamma^c_{\nu;i}\,G_\nu\,T_{\nu\mu}^\npo \ , \non
\Gamma_{\mu;i}^a	&=&\gamma^a_{\mu;i}+\sum_\nu  T_{\mu\nu}^\npo\,G_\nu\,\gamma^a_{\nu;i} \ , \non
\Sigma_{ij}		&=&\sum_\mu  \gamma^c_{\mu;i}\,G_\mu\,\Gamma^a_{j;\mu} \;,
\label{dressed}
\end{eqnarray}
where $\Sigma$ is the self-energy. The indices $i$ and $j$ label the $s$-channel states or a contact diagram in a given partial wave. 

The pole part is then given by
\be
T^{\po}_{\mu\nu}=\Gamma^a_{\mu;i}\, D_{ij} \, \Gamma^c_{\nu;j}
\label{tpo}
\ee
with the resonance propagator $D_{ij}$. 
For example, if there are two $s$-channel resonances with masses $m_1$ and $m_2$ (indices $i,j\in \{1,2\}$) plus one contact term
(indices $i,j=3$), the expression reads
\begin{eqnarray}
\Gamma^a_\mu=(\Gamma^a_{\mu;1},\Gamma^a_{\mu;2},\Gamma^a_{\mu;3}), \quad
\Gamma^c_\mu=\left(
\begin{matrix}
\Gamma^c_{\mu;1}\\
\Gamma^c_{\mu;2}\\
\Gamma^c_{\mu;3}
\end{matrix}
\right), \non
D^{-1}=\left(\begin{matrix}
E-m^b_1-\Sigma_{11}&&-\Sigma_{12}&&-\Sigma_{13}\\
-\Sigma_{21}     &&E-m^b_2-\Sigma_{22}&&-\Sigma_{23}\\
-\Sigma_{3 1}&&-\Sigma_{32} &&m_N-\Sigma_{33} 
\end{matrix}
\right). \non
\label{3res}
\end{eqnarray}
The decomposition into pole and non-pole part performed here has mostly technical reasons: The numerical evaluation of the non-pole part is
much more time-consuming than the evaluation of the pole part. This leads to an effective, nested fitting workflow as discussed in
detail in Refs.~\cite{Ronchen:2012eg,Ronchen:2014cna}. Here, we have introduced the contact terms technically on the same footing as
resonances, which allows to fit those terms computationally more effectively.

In Sec.~2.2 of Ref.~\cite{Ronchen:2012eg}, the renormalization of the nucleon mass and coupling in the presence of two bare $s$-channel states in the $P_{11}$ 
partial wave was derived. With the implementation of contact diagrams, this procedure has to be extended to two resonances and one contact diagram.
This is addressed in Appendix~\ref{renorma3}.

\subsection{Photon-induced reactions}
\label{formalism_photo}

A field-theoretical description of the photoproduction amplitude within a gauge-invariant framework that respects the generalized off-shell Ward-Takahashi identity~\cite{Haberzettl:2011zr,  Haberzettl:2006bn, Haberzettl:1997jg}   was successfully applied in the analysis of pion photoproduction in Ref.~\cite{Huang:2011as}. In this study an earlier version of the J\"ulich model was utilized to provide the hadronic final-state interaction. This field-theoretical method allows one to gain insight into the microscopic reaction dynamics of the photo-interaction. By contrast, the semi-phenomenological approach to pseudoscalar meson photoproduction, developed in Ref.~\cite{Ronchen:2014cna} and used here, is more flexible and facilitates the analysis of a large amount of data. Here, the photo-interaction kernel is approximated by energy-dependent polynomials, while the hadronic final-state interaction is provided by the J\"ulich DCC model described in Sec.~\ref{formalism_pion}. The formalism is inspired by the GW-SAID CM12 parameterization~\cite{Workman:2012jf} and will be applied in the present study.
Nonetheless, we consider the present analysis as an intermediate step towards an expansion of the field-theoretical framework of Ref.~\cite{Huang:2011as}.
A detailed introduction to the semi-phenomenological approach was given in Sec.~2.2 of Ref.~\cite{Ronchen:2014cna}. In the following, we recapitulate the basic elements.

The multipole amplitude of the photoproduction process is given by 
\begin{multline}
M_{\mu\gamma}(q,E)=V_{\mu\gamma}(q,E) \\+\sum_{\kappa}%\int\limits_0^\infty dp\,p^2\,
T_{\mu\kappa}(q,p,E)G^{}_\kappa(p,E)V_{\kappa\gamma}(p,E)\ ,
\label{m2}
\end{multline}
where the index $\gamma$ denotes the initial $\gamma N$ state. $T_{\mu\kappa}$ is the hadronic half-off-shell $T$-matrix introduced in Sec.~\ref{formalism_pion} with the intermediate (final) meson-baryon channel $\kappa$  ($\mu$) and the corresponding off-shell momentum $p$ (on-shell momentum $q$). Integration over the intermediate off-shell momentum $p$
similar to Eq.~(\ref{scattering}) is suppressed here in the notation, following the convention
of Eq.~(\ref{deco2}). In the present analysis the photon is allowed to couple to the intermediate channels $\kappa=\pi N$, $\eta N$ and $\pi\Delta$, while we have $\pi N$ and $\eta N$ as final states $\mu$.

The photoproduction kernel $V_{\mu\gamma}$ is written as 
\be
V_{\mu\gamma}(p,E)=\alpha^\npo_{\mu\gamma}(p,E)+\sum_{i} \frac{\gamma^a_{\mu;i}(p)\,\gamma^c_{\gamma;i}(E)}{E-m_i^b} \ .
\label{vg}
\ee
Here, $\alpha^\npo_{\mu\gamma}$ stands for the photon coupling to the non-pole part of the photoproduction kernel. The tree-level coupling of the $\gamma N$ channel to the  nucleon and $\Delta$ resonances is represented by the vertex function $\gamma^c_{\gamma;i}$, where $i$ denotes the resonance number in a given partial wave. The hadronic resonance annihilation vertex $\gamma^a_{\mu;i}$ is exactly the same as in Eq.~(\ref{vblubb}), which results in the cancellation of the explicit singularity at $E=m^b_i$. 
Both quantities, $\alpha^\npo_{\mu\gamma}$ and $\gamma^c_{\gamma;i}$, are approximated by energy-dependent polynomials $P$,
\begin{eqnarray}
\alpha^\npo_{\mu\gamma}(p,E)&=& \frac{ \tilde{\gamma}^a_{\mu}(p)}{\sqrt{m_N}} P^{\text{NP}}_\mu(E)  \nonumber \\
\gamma^c_{\gamma;i}(E)&=& \sqrt{m^{}_N} P^{\text P}_i(E) \ .
\label{vg_poly}
\end{eqnarray} 
In Eq.~(\ref{vg_poly}), the vertex function $\tilde{\gamma}^a_{\mu}$ has the same form as ${\gamma}^a_{\mu,i}$ in Eq.~(\ref{vg}) but without any dependence on the resonance number $i$. The polynomials $P$, for a given multipole, are parameterized as
\ba
P^{\text P}_i(E)&=&  \sum_{j=1}^{\ell_i}  g^{\text P}_{i,j} \left( \frac{E - E_s}{m_N} \right)^j
 e^{-\lambda^\po_{i}(E-E_s)} 
\non
P^{\text{NP}}_\mu(E) &=&  \sum_{j=0}^{\ell_{\mu}} g^{\text{NP}}_{\mu,j} \left( \frac{E - E_s}{m_N}\right)^j
 e^{-\lambda^{\text{NP}}_{\mu}(E-E_s)} \ ,\non
\label{polys}
\ea
with $g$ and $\lambda>0$ being free parameters that are fitted to the data. The upper limits of the summation indices $j$, $\ell_i$ and $\ell_{\mu}$, are chosen  so as to permit a good data description, but are restricted to be less than 4. In order to fulfill the decoupling theorem, which states that resonance contributions are parametrically suppressed at threshold, the summation for $P^{\text P}$ starts with $j=1$. The expansion point $E_s$ is chosen to be close to the $\pi N$ threshold, $E_s=1077$~MeV. In this way, the factor $e^{-\lambda(E-E_s)}$ absorbs the potentially strong energy dependence at the $\gamma N$ threshold, which is not too far from the $\pi N$ threshold. Moreover, this factor guarantees a well-behaved multipole amplitude in the high-energy limit, although a more quantitative matching to Regge amplitudes remains to be done~\cite{Huang:2008nr}.

In order to achieve a good description of the high-precision data for pion photoproduction close to threshold, we take into account some isospin breaking effects, i.e. we apply different threshold energies for the $\pi^0p$ and the $\pi^+ n$ channels, as explained in Sec.~2.3 in Ref.~\cite{Ronchen:2014cna}. Similarly, we take the physical threshold of the $\eta p$ final state when calculating observables. Note that, in general, isospin-averaged masses are used in the J\"ulich model. 

A multipole decomposition of the photoproduction amplitude of pseudoscalar mesons can be found in Appendix~A of Ref.~\cite{Ronchen:2014cna}. 
%%%%%%%%%%%%%%%%%%%%%%%%%%%%%%%%%%%%%%%%%%%%
\section{Results}

\subsection{Database and free parameters}
\label{sec:data_fit}

The database of the present study comprises  the hadronic data used in Ref.~\cite{Ronchen:2012eg}. This represents the world database on the reactions $\pi N\to\eta N, K\Lambda$, and $K\Sigma$ up to $E\sim 2.3$~GeV, plus the $\pi N\to\pi N$ WI08 energy-dependent solution of the GWU/INS SAID group~\cite{Workman:2012hx}. In addition, we include almost all published data on pion photoproduction off the proton up to $E\sim 2.3$~GeV, with some forward regions at high energy excluded~\cite{Ronchen:2014cna}. Third, in this study we add the world database for the reaction $\gamma p\to \eta p$, again up to $E\sim 2.3$~GeV. These data were taken from the GW-SAID database~\cite{Workman:2012hx}.

\begin{table}
\begin{center}
\renewcommand{\arraystretch}{1.6}
\begin {tabular}{l|cc} 
\hline\hline
& \multicolumn{1}{c}{Fit {A}} & \multicolumn{1}{c}{Fit {B}}\\ \hline
& \includegraphics[width=1.5cm]{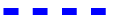}	&\includegraphics[width=1.5cm]{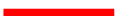}\\
$\pi N\to\pi N$ & \multicolumn{2}{c}{PWA  GW-SAID WI08 \cite{Workman:2012hx} }\\
$\pi^-p\to\;\eta n$ &\multicolumn{2}{c}{$d\sigma/d\Omega$, $P$ }\\
$\pi^-p\to\; K^0 \Lambda$ &\multicolumn{2}{c}{$d\sigma/d\Omega$, $P$, $\beta$ }\\
$\pi^-p\to\; K^0 \Sigma^0$ &\multicolumn{2}{c}{$d\sigma/d\Omega$, $P$ }\\
$\pi^-p\to\; K^+ \Sigma^-$ &\multicolumn{2}{c}{$d\sigma/d\Omega$}\\
$\pi^+p\to\; K^+ \Sigma^+$ &\multicolumn{2}{c}{$d\sigma/d\Omega$, $P$ , $\beta$}\\ 
&\multicolumn{2}{c}{$\sim$ {6000 data points}}\\ \hline
$\gamma p\to \pi^0p$ & \multicolumn{2}{c}{$d\sigma/d\Omega$, $\Sigma$, $P$, $T$, $\Delta\sigma_{31}$, $G$, $H$ }\\
$\gamma p\to \pi^+n$ & \multicolumn{2}{c}{$d\sigma/d\Omega$, $\Sigma$, $P$, $T$, $\Delta\sigma_{31}$, $G$, $H$ }\\
$\gamma p\to \eta p$ & \multicolumn{1}{l}{$d\sigma/d\Omega$, $P$, $\Sigma$ }&\multicolumn{1}{|l}{$d\sigma/d\Omega$, $P$, $\Sigma$, {$T$, $F$} }\\
& {29,392 data points} &\multicolumn{1}{|c}{{29,680 data points}}\\
\hline\hline
\end {tabular}
\end{center}
\caption{Data included in fits A and B. In contrast to fit A, fit B contains the recent MAMI measurements of $T$ and $F$~\cite{Akondi:2014ttg}.}
\label{tab:fit_charac}
\end{table}

	In a first fit, called fit A in the following, we include all data except the recent eta photoproduction measurement of the 
transverse target asymmetry $T$ and the beam-target asymmetry $F$ by the A2 collaboration at MAMI~\cite{Akondi:2014ttg}. These data are added in a second fit, fit B. Performing these two fits allows us to estimate the influence of the new polarization data on the extracted resonance spectrum. An overview of the data included in the fits can be found in Table~\ref{tab:fit_charac}.

Compared to the high-precision data nowadays available in case of pseudoscalar meson photoproduction, the data situation for the pion-induced reactions is difficult in large parts, the lack of polarization measurements being one of the major issues. In Sec.~3 of Ref.~\cite{Ronchen:2012eg} the situation for the individual hadronic channels was discussed in detail. In the present study, we adopt the systematic errors and mainly also the special weights applied to certain data sets.

Moreover, we continue along the lines of Ref.~\cite{Ronchen:2014cna} and apply an additional systematic error of 5~\% to all photon-induced data in order to account for discrepancies in the data. The amount of 5\% is an estimate at this point. More advanced techniques have been applied, e.g., by the GWU/SAID group allowing for normalization corrections~\cite{Arndt:2006bf, Workman:2012jf}. We plan to
improve our analysis along these lines in the future. To compensate for the smaller number of data points for $T$ and $F$ in eta photoproduction, those data are weighted in fit B with generic factors around 10 as found necessary for obtaining
satisfactory fit results. To achieve a good description of the data at higher energies, additional weights have to be applied. The situation is similar in case of the beam asymmetry $\Sigma$ in $\gamma p\to\eta p$. Compared to the number of data points for the differential cross section, 5680, only a few are available for $\Sigma$, namely 189. Furthermore, we did not attempt to achieve a good description of the recoil polarization $P$ for $\gamma p\to\eta p$ as only seven data points are available and their influence on the fit is very limited. The data are, however, included in the fit but no special weights were applied. 

In the J\"ulich approach the free parameters tied to the hadronic interaction are the bare coupling constants and masses of the $s$-channel resonances, the strengths of the contact terms, and the cut-off parameters in the $t$- and $u$-channel diagrams constituting $T^{\text{NP}}$. In addition, there are certain couplings in some of the latter diagrams that cannot be connected to other coupling constants via SU(3) flavor symmetry and have to be fitted to data as well. In the present study we do not alter the $T^{\text{NP}}$ parameters but employ the values found in the J\"ulich2012 analysis of pion-induced reactions~\cite{Ronchen:2012eg}. 

We do, however, vary the parameters tied to hadronic resonances and contact terms. This is a necessity in the current situation, given that the data of pion-induced
eta production is of less quality than the eta photoproduction data. The higher-quality photoproduction data can, thus,
help to constrain the hadronic amplitude.
In both fits A and B the number of $s$-channel resonance parameters amounts to 128. For each of the 11 genuine $I=1/2$ and 10 genuine $I=3/2$ resonances (i.e., those resonances included in form of a pole diagram in the potential~\cite{Doring:2010ap, Ronchen:2012eg}) the parameters are given by one bare mass and the couplings to the channels $\pi N$, $\rho N$, $\eta N$, $\pi\Delta$, $K\Lambda$ and $K\Sigma$ as allowed by isospin. Contact terms introduced in Eq.~(\ref{ctct}) are switched on in both fits in the $S_{11}$ and $P_{11}$ partial waves for the $\pi N$ and $\eta N$ channel, and in the $P_{13}$ partial wave for the $\pi N$, $\eta N$, $\pi\Delta$, $K\Lambda$ and $K\Sigma$ channel, giving rise to 9 additional fit parameters.
The free parameters that are used to tune the interaction of the photon with hadrons are the resonance parameters $g^\po_{i,j}$ and $\lambda^\po_{i}$  and the non-pole couplings $g^\npo_{\mu,j}$ and $\lambda^\npo_{\mu,j}$ with $\mu=\pi N$, $\pi\Delta$ or $\eta N$, cf. Eq.~(\ref{polys}). Formally, all fit parameters up to $j=3$ are implemented in the computer code. However, the actual number of fit parameters is chosen as required by data and the remaining parameters are set to zero. In order to achieve a good description of $T$ and $F$ in fit B, additional fit parameters that were set to zero in fit A had to be released. Thus,  we have 443 parameters tied to the photon interaction in fit A and 456 in fit B. 
In total, the number of fit parameters adds up to 580 in fit A and 593 in fit B.

The free parameters are adjusted to the data in simultaneous fits of all pion- and photon-induced reactions using MINUIT on the JUROPA supercomputer at the Forschungszentrum J\"ulich.

\subsection{Fit results}
\label{sec:fit}

In the following, only selected fit results for the reactions $\gamma p\to \eta p$ and $\pi^-p\to\eta n$ are shown. 
Data sets with energies that differ by less than 5.5~MeV are sometimes displayed in the same graph.
The full fit results for all pion- and photon-induced reactions included in this analysis can be found online~\cite{Juelichmodel:online}. 

The definition of the various photoproduction observables in terms of the multipole amplitudes $M_{\mu\gamma}$ is given in Appendix~B of Ref.~\cite{Ronchen:2014cna}. The convention agrees with the one of the SAID group~\cite{Sandorfi:2011nv, Arndt:2002xv}.

\subsubsection{$\gamma p\to\eta p$}

In Figs.~\ref{fig:dsdoetap1} and \ref{fig:dsdoetap2} we show selected results for the differential cross section. In the threshold region only very small 
differences can be observed between fits A and B. Starting at $E \sim1600$~MeV, the two fits show some discrepancies at extreme angles, fit A providing a
 slightly better description of the data at medium energies, cf. $E=1629$~MeV in Fig.~\ref{fig:dsdoetap1}. At higher energies, the differences are most 
apparent at very forward angles. However, the data situation does not allow for an assessment regarding which of the fits is best. This can be seen, e.g.,
at $E=2006$~MeV in Fig.~\ref{fig:dsdoetap2} where both fits seem to describe the data equally well.

\setlength{\unitlength}{\textwidth}

\begin{figure*}%[htbp]
\begin{center}
\begin{picture}(1,0.75)
\put(0.03,0.0){
\includegraphics[width=0.96\textwidth]{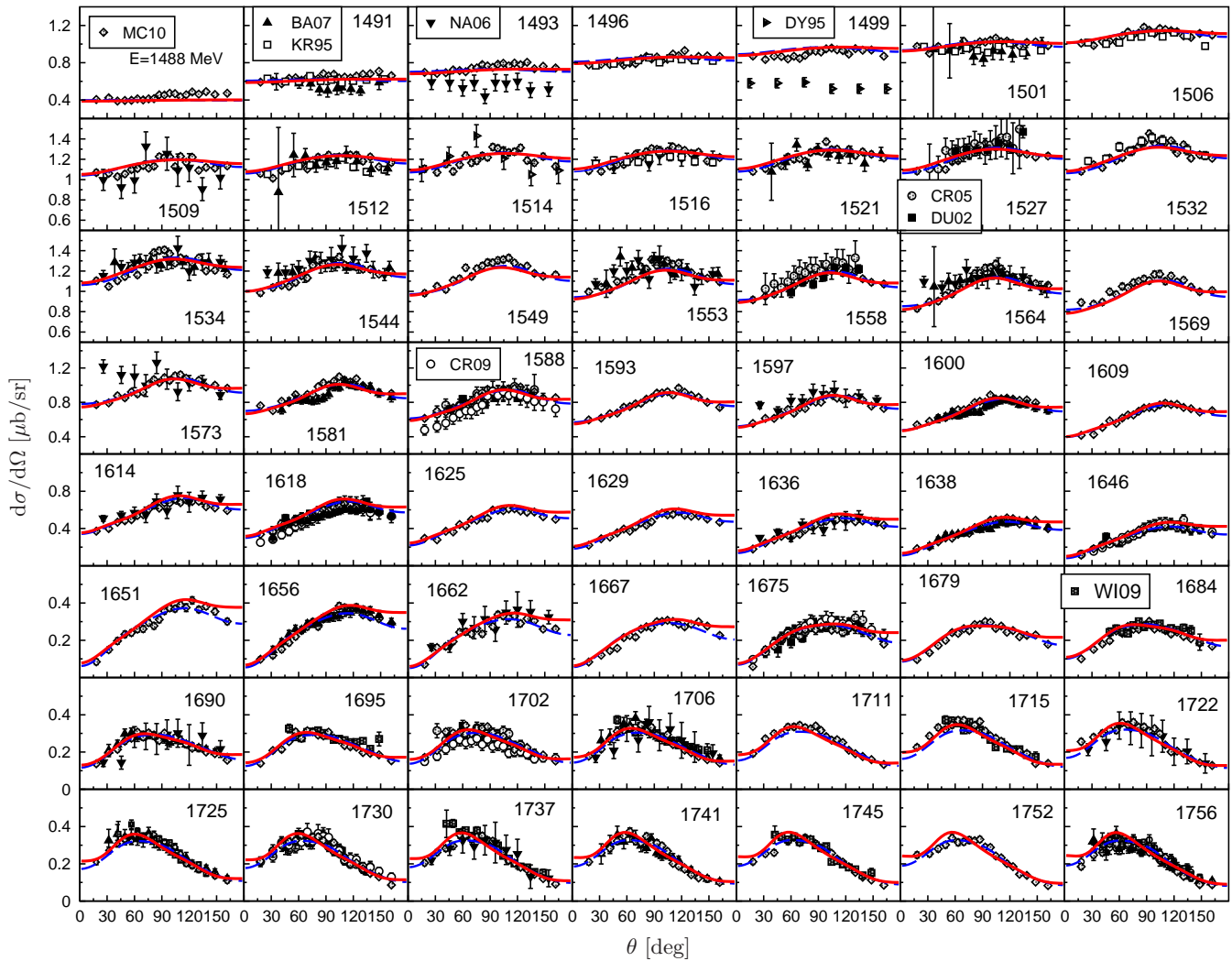}
}
\put(0,0.33){
\begin{turn}{90}
d$\sigma/$d$\Omega$ [$\mu$b/sr]
\end{turn}
}
\put(0.5,-0.024){
$\theta$ [deg]
}
\end{picture}

\end{center}
\caption{Differential cross section of the reaction $\gamma p\to \eta p$. Dashed (blue) line: fit A; solid
(red) line: fit B; data: MC10~\cite{McNicoll:2010qk}, BA07~\cite{Bartalini:2007fg}, KR95~\cite{Krusche:1995nv}, NA06~\cite{Nakabayashi:2006ut}, DY95~\cite{Dytman:1995vm}, CR05~\cite{Crede:2003ax}, DU02~\cite{Dugger:2002ft}, CR09~\cite{Crede:2009zzb}, WI09~\cite{Williams:2009yj}. In this and all following figures, the numbers in the plots specify the pertinent center-of-mass energy $E$[MeV].}
\label{fig:dsdoetap1}
\end{figure*}

\begin{figure*}
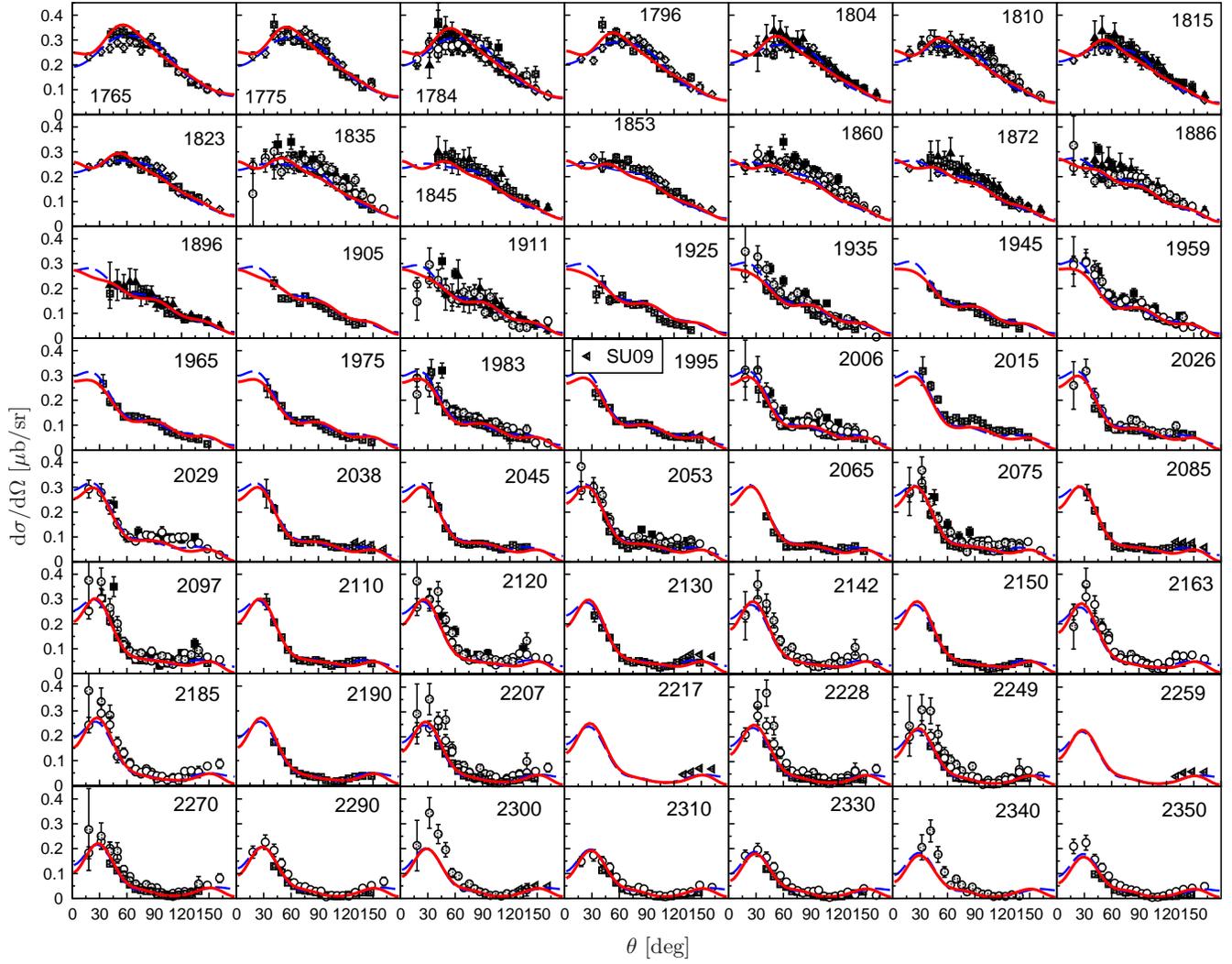
%[htbp]
\begin{center}

\begin{picture}(1,0.84)
\put(0.03,0.2){\includegraphics[width=0.96\textwidth]{dsdo_etap_2.eps}
}
\put(0.03,0.095){\includegraphics[width=0.96\textwidth]{dsdo_etap_3.eps}
}
\put(0,0.4){
\begin{turn}{90}
d$\sigma/$d$\Omega$ [$\mu$b/sr]
\end{turn}
}
\put(0.5,0.07){
$\theta$ [deg]
}
\end{picture}
\vspace{-1.9cm}
\end{center}
\caption{Differential cross section of the reaction $\gamma p\to \eta p$. Dashed (blue) line: fit A; solid
(red) line: fit B; data: cf. Fig.~\ref{fig:dsdoetap1} and SU09~\cite{Sumihama:2009gf}. }
\label{fig:dsdoetap2}
\end{figure*}

  %------------------------------

\begin{figure*}%[htbp]
\begin{center}
\begin{picture}(1,0.34)
\put(0.03,0.03){
\includegraphics[width=0.96\textwidth]{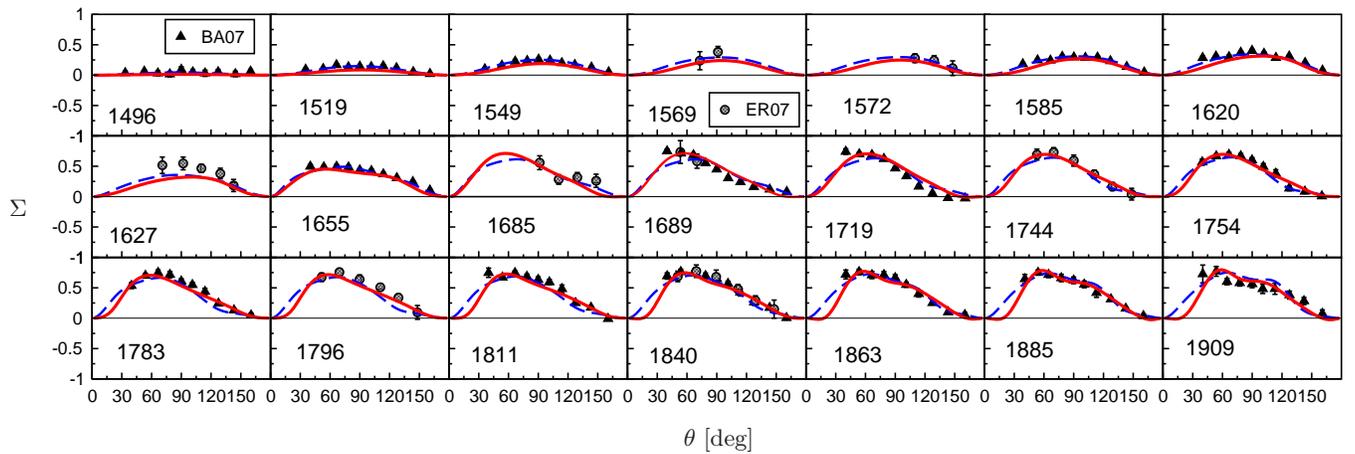} 
}
\put(0,0.17){
$\Sigma$
}
\put(0.5,0){
$\theta$ [deg]
}
\end{picture}

\end{center}
\caption{Beam asymmetry of the reaction $\gamma p\to \eta p$. Dashed (blue) line: fit A; solid
(red) line: fit B; data: BA07~\cite{Bartalini:2007fg}, ER07~\cite{Elsner:2007hm}.}
\label{fig:setap1}
\end{figure*}

The situation is similar in case of the beam asymmetry $\Sigma$ in Fig.~\ref{fig:setap1}. Disagreements of the fits A and B are visible predominantly at higher energies and at forward angles where there are no data.
  %------------------------------
\begin{figure}%[htbp]
\begin{center}
\includegraphics[width=1\linewidth]{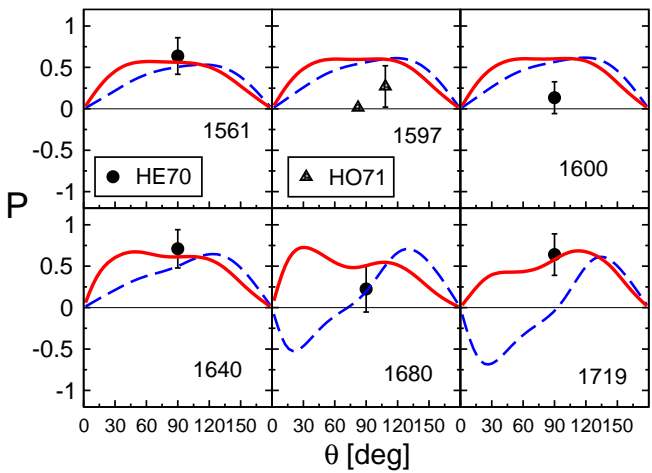} 
\end{center}
\caption{Recoil polarization of the reaction $\gamma p\to \eta p$. Dashed (blue) line: fit A; solid
(red) line: fit B; data: HE70~\cite{Heusch:1970tr}, HO71~\cite{Hongoh:1971pv}.}
\label{fig:polaetap}
\end{figure}

Only seven data points are available for the recoil polarization $P$, cf. Fig.~\ref{fig:polaetap}. Although the results of our two fits are very different, especially at higher energies, both describe the data more or less well but with some deficiencies. A larger database of this observable may help constrain the partial wave content.
%--------------------------

  %------------------------------
\begin{figure*}
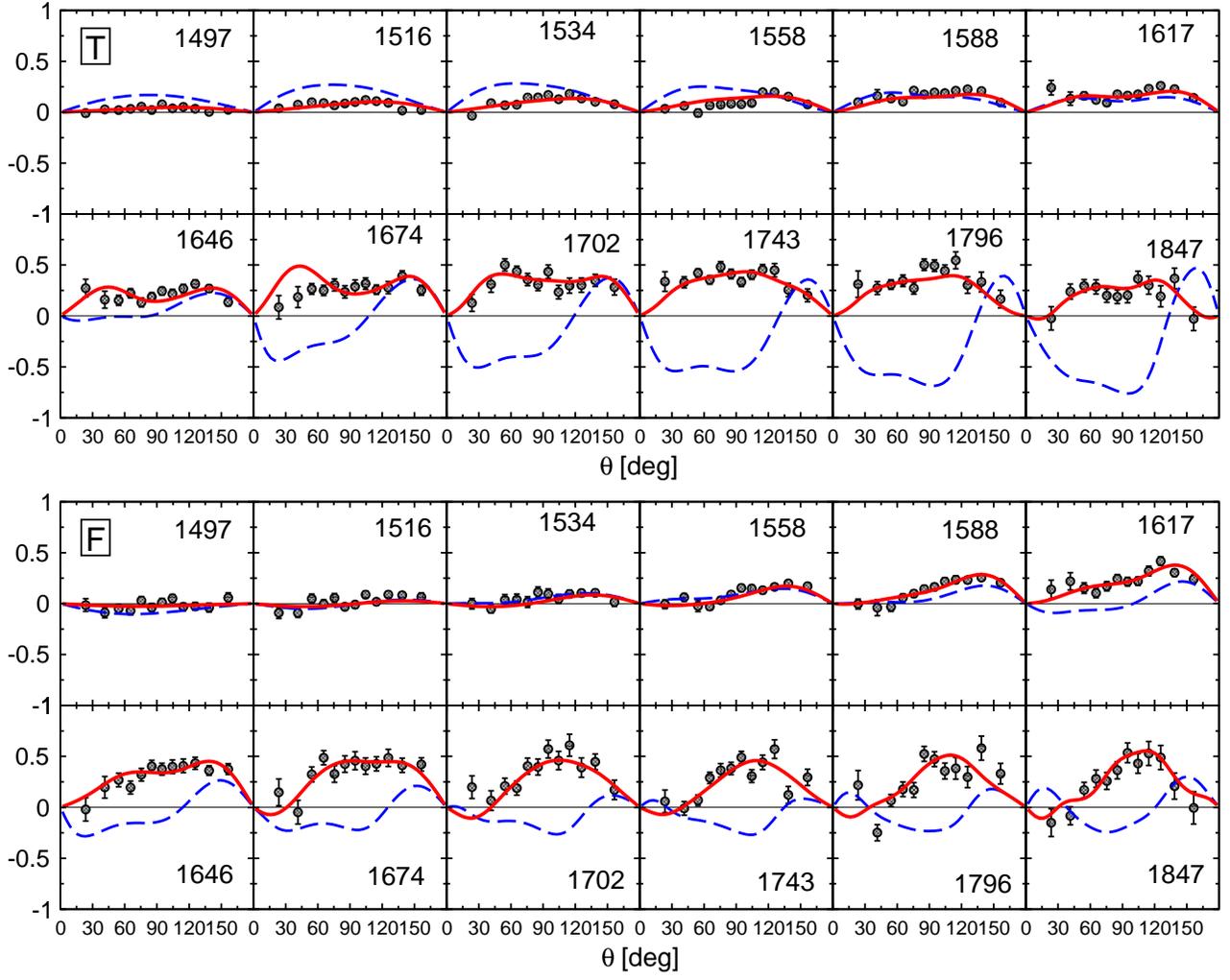
%[htbp]
\begin{center}
\begin{picture}(1,0.77)
\put(0.018,0.4){
\includegraphics[width=0.94\linewidth]{T_etap_MAMI_A2_paper.eps} 
}
\put(0.018,0.02){
\includegraphics[width=0.94\linewidth]{F_etap_MAMI_A2_paper.eps} 
}
\end{picture}
\end{center}
\vspace{-1cm}
\caption{Transverse target asymmetry $T$ and beam-target asymmetry $F$ in the reaction $\gamma p\to \eta p$. Dashed (blue) line: prediction of fit A; solid
(red) line: fit B; data: Ref.~\cite{Akondi:2014ttg}.}
\label{fig:TFetap}
\end{figure*}

The fit results for the transverse target asymmetry $T$ and the beam-target asymmetry $F$ can be found in Fig.~\ref{fig:TFetap}. These data were only included in fit B, meaning that the blue dashed line in Fig.~\ref{fig:TFetap} (fit A) represents a prediction for these observables. Note that older data for $T$ from Ref.~\cite{Bock:1998rk}, that are partially in conflict with the MAMI data, were not fitted. The prediction of fit A at lower energies $E < 1600$~MeV is acceptable for $T$ and good for $F$, while the shortcomings of the prediction are considerable at higher energies. However, once the data are included (fit B) a good description over the whole energy range is achieved.

%--------------------------

\subsubsection{$\pi^-p\to\eta n$}

\begin{figure}[htbp]
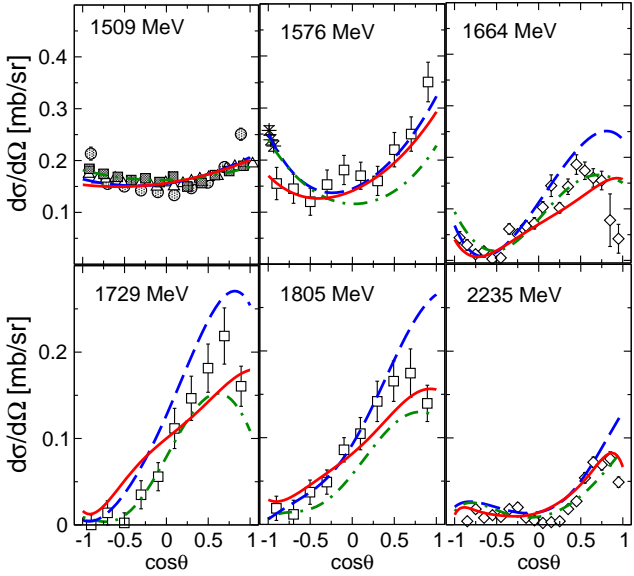

\begin{center}
\includegraphics[width=0.393\linewidth]{dsdo_etaN_1509.eps} \hspace{-0.26cm}
\includegraphics[width=0.292\linewidth]{dsdo_etaN_1576.eps} \hspace{-0.26cm} 
\includegraphics[width=0.292\linewidth]{dsdo_etaN_1664.eps} \\ \vspace{-0.1cm}
\includegraphics[width=0.393\linewidth]{dsdo_etaN_1729.eps} \hspace{-0.26cm}
\includegraphics[width=0.292\linewidth]{dsdo_etaN_1805.eps} \hspace{-0.26cm} 
\includegraphics[width=0.292\linewidth]{dsdo_etaN_2235.eps} 
\end{center}
\caption{Differential cross section of the reaction $\pi^-p\to\eta n$. Dashed (blue) line: fit A; solid
(red) line: fit B; dash-dotted (green) line: fit A$_{\text{had}}$ of Ref.~\cite{Ronchen:2012eg}. Data: filled circles from Ref.~\cite{Bayadilov:2008zz};
filled squares from Ref.~\cite{Prakhov:2005qb}; empty triangles up from Ref.~\cite{Kozlenko:2003hu}; stars from Ref.~\cite{Debenham:1975bj}; empty squares from Ref.~\cite{Richards:1970cy};
empty diamonds from Ref.~\cite{Brown:1979ii}. Note that the data situation for this reaction is problematic, see text. }
\label{fig:etaN_dsdo}
\end{figure}

In Fig.~\ref{fig:etaN_dsdo} we show selected fit results for the differential cross section of the reaction $\pi^-p\to\eta n$. In addition to fits A and B, we display results from the old fit A of the J\"ulich2012 analysis~\cite{Ronchen:2012eg}, in which only pion-induced reactions were analyzed. We call the latter fit A$_{\rm had}$ in the following, where the subscript serves as a reminder that the fit in Ref.~\cite{Ronchen:2012eg} included only hadronic data. Since the resonance vertex function $\gamma^a_{\eta N}$ appears in the construction of the hadronic amplitude in Eq.~(\ref{vblubb}) and also in the photoproduction kernel in Eq.~(\ref{vg}), the bare parameters of  $\gamma^a_{\eta N}$ have influence on the pion- as well as on the photon-induced production of the $\eta N$ final state. 
As much less data are available for $\pi^-p\to\eta n$ compared to $\gamma p\to \eta p$, the photon-induced eta production does impose constraints on the fit results of the hadronic $\eta N$ channel.

As can be seen in Fig.~\ref{fig:etaN_dsdo}, starting already at energies of $E\sim1.5$~GeV, data sets from different experimental groups show conflicting behavior. Moreover, at energies $E>1800$~MeV all data for the differential cross section of $\pi^-p\to\eta n$ stem from a measurement~\cite{Brown:1979ii} deemed problematic due to a miscalibration of the beam momentum, see Ref.~\cite{Clajus:1992} for details. Those data enter the fit with a much reduced weight and all three fits yield different results. This is most apparent at higher energies, where fits A and B are dominated by photon-induced data. 
Here, fits A and B mostly differ at extreme forward angles while for other angles they coincide reasonably well. This discrepancy
has been also noted for the differential cross section and beam asymmetry in fits A and B in eta photoproduction, mostly due to less precise data at forward angles. 

\begin{figure}[htbp]
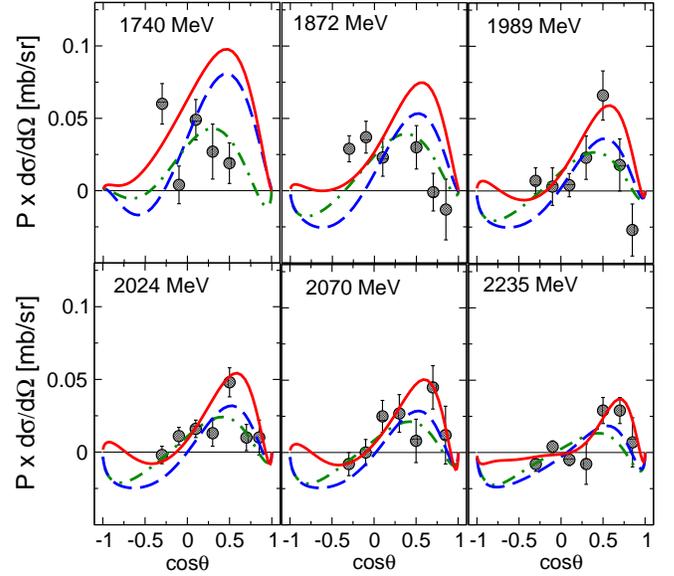

\begin{center}
\includegraphics[width=0.412\linewidth]{pola_x_dsdo_etaN_1740.eps} \hspace{-0.26cm}
\includegraphics[width=0.292\linewidth]{pola_x_dsdo_etaN_1872.eps} \hspace{-0.26cm} 
\includegraphics[width=0.292\linewidth]{pola_x_dsdo_etaN_1989.eps} \\ \vspace{-0.1cm}
\includegraphics[width=0.412\linewidth]{pola_x_dsdo_etaN_2024.eps} \hspace{-0.26cm}
\includegraphics[width=0.292\linewidth]{pola_x_dsdo_etaN_2070.eps} \hspace{-0.26cm} 
\includegraphics[width=0.292\linewidth]{pola_x_dsdo_etaN_2235.eps} 
\end{center}
\caption{Polarization of the reaction $\pi^-p\to\eta n$. Dashed (blue) line: fit A; solid
(red) line: fit B; dash-dotted (green) line: fit A$_{\text{had}}$ of Ref.~\cite{Ronchen:2012eg}. Data: Ref.~\cite{Baker:1979aw}. Note that the data situation for this reaction is problematic, see text. }
\label{fig:etaN_pola}
\end{figure}

In Fig.~\ref{fig:etaN_pola} we present results of the recoil polarization in $\pi^-p\to\eta n$.  In principle, the only published data for this observable~\cite{Baker:1979aw} exhibit the same problems as the differential cross sections of Ref.~\cite{Brown:1979ii} because the same experimental set up was used. Those data are fitted with a very low weight, too. As in the case of the differential cross section, the differences in the three fits are obvious. However, in view of the discussed quality issues
with the available data for $\pi^-p\to\eta n$, we see no reason to enforce a better data description. 

The comparison made here demonstrates the need for a pion beam to re-measure the reaction $\pi^-p\to\eta n$. Note that only four observables are needed for a complete experiment, as discussed in the Introduction.

\subsection{Multipoles}

\begin{figure*}%[htbp]
\begin{center}
\includegraphics[width=1\textwidth]{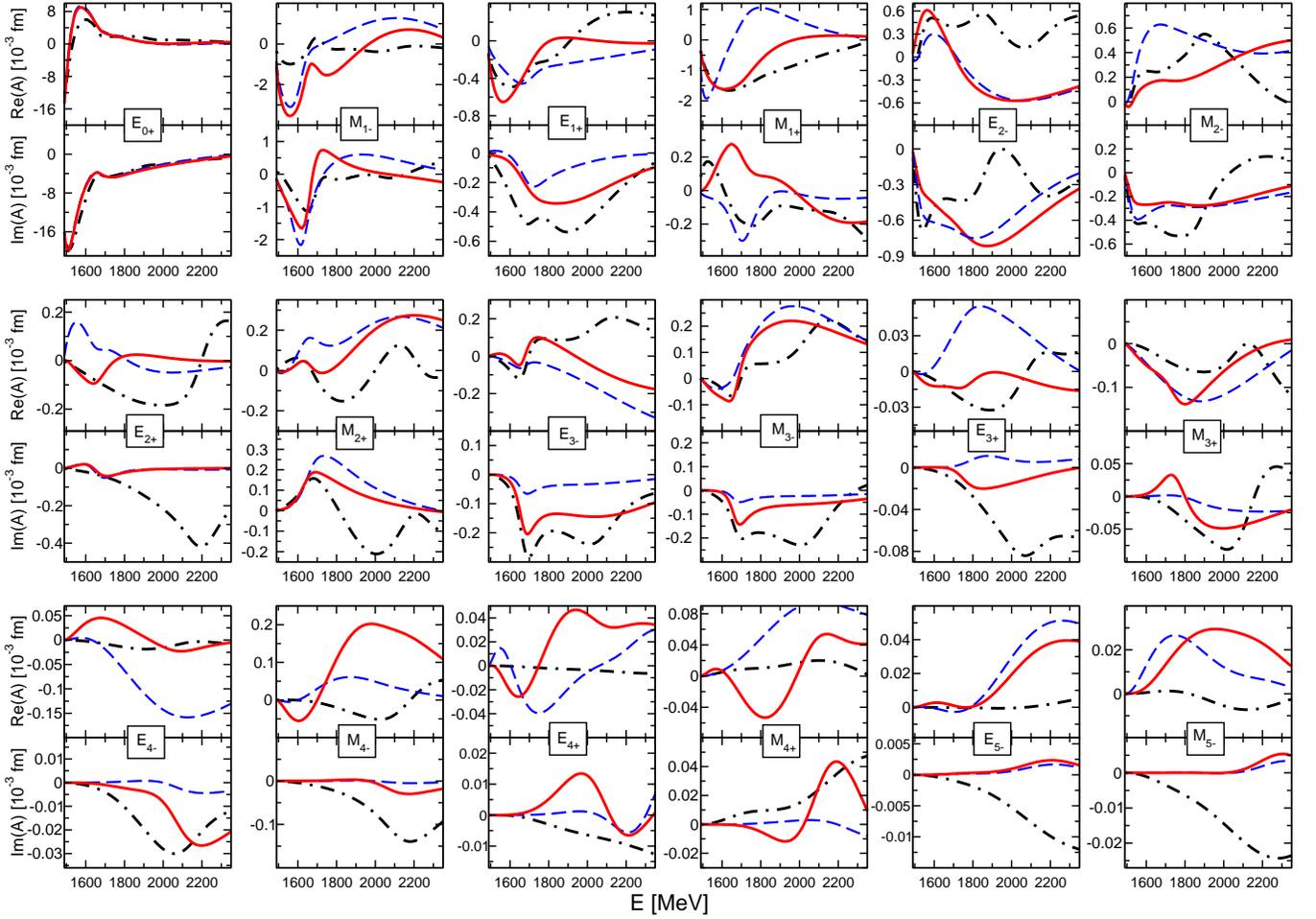}
\end{center}
\caption{Electric and magnetic multipoles for the reaction $\gamma p\to \eta p$. (Black) dash-dotted line: BG 2014-02 solution \cite{Gutz:2014wit}. Dashed (blue) line:
fit A; solid (red) line: fit B.}
\label{fig:mltp_etap}
\end{figure*}

In Fig.~\ref{fig:mltp_etap} the electric and magnetic multipoles from fits A and B for the reaction $\gamma p\to \eta p$ are shown. Additionally, we display the multipoles of the Bonn-Gatchina BG2014-02 solution~\cite{Gutz:2014wit}. Note that those amplitudes were not included in the fit.

Generally speaking, the multipoles extracted in our analysis and the ones of the Bonn-Gatchina group exhibit large differences. An exception is the $E_{0+}$ and, to a certain degree, also the $M_{1-}$ multipole. While for $E_{0+}$ differences between our fits A and B are hardly noticeable at all, the $M_{1-}$ multipole features clearly visible deviations in the two fits at energies higher than $E\sim$1650~MeV. As will be discussed in Sec.~\ref{spec_res}, this is the energy regime where the pole of the $N(1710)1/2^+$ resonance is located.
Among the lower multipoles, the $M_{1+}$ shows the most striking discrepancies between fits A and B, see also the discussion on the influence of variations in the $P_{13}$ partial wave on the description of $T$ and $F$ in Sec.~\ref{spec_res}. For higher multipoles fit B sometimes shows a stronger energy dependence than fit A, cf.  $E_{4-}$, $M_{4-}$ and $E_{4+}$, $M_{4+}$. In summary, the new MAMI data for $T$ and $F$ have a large impact on the multipole amplitudes.

We observe that for lower partial waves, the eta photoproduction multipoles of the present study exhibit less agreement with the Bonn-Gatchina multipoles than our pion photoproduction multipoles with the Bonn-Gatchina BG2014-02~\cite{Gutz:2014wit} or the GW-SAID CM12 solution~\cite{Workman:2012jf}. This suggests that the multipole content of the reaction $\gamma p\to\eta p$ is much less established than in the case of pion
photoproduction where the various analyses agree better. Figures showing the pion photoproduction multipoles can be found online~\cite{Juelichmodel:online}.

%%%%%%%%%%%%%%%%%%%%%%%%%%%%%%%

\section{Resonance spectrum}
\label{resonances}

A resonance state is uniquely defined by its pole position in the complex energy plane, the residues associated with the channel transitions, and the Riemann sheet the pole is located on.
With the exception of the physical sheet of the lowest lying channel, the poles can appear on various Riemann sheets, but not all of them
 are of physical interest. Usually, only the poles on the sheet which is closest to the physical axis are considered. We select this sheet by
 rotating the right-hand cuts of all channels in the direction of the negative imaginary $E$ axis. In this way, we define the {\it second sheet} 
 where all poles extracted in the present study lie. See Ref.~\cite{Doring:2009yv} for a detailed discussion.

In order to determine the pole positions, the scattering amplitude has to be continued to the second Riemann sheet.  For this purpose we
 apply the method of analytic continuation following Ref.~\cite{Doring:2009yv} where the amplitude on the second sheet is accessed via a
 contour deformation of the momentum integration. The calculation of the residues proceeds via the formalism illustrated in Appendix~C of Ref.~\cite{Doring:2010ap}. Definitions of the {normalized residue} and the branching ratio into a specific channel are given in Sec.~4.1 of
 Ref.~\cite{Ronchen:2012eg}. In case of the latter two quantities we use the same definitions as the Particle Data Group \cite{Agashe:2014kda}.
For a reliable extraction of the resonance parameters, the correct structure of branch points, including the complex branch points of the
channels including unstable particles $\pi\Delta$, $\sigma N$ and $\rho N$, is crucial. In Ref.~\cite{Ceci:2011ae} it was shown that the absence of the 
 latter might lead to false resonance signals. 

The definition of the photocouplings at the pole $\tilde A^{h}_{\rm pole}$,
\be
\tilde A^h_{\rm pole}=A^h_{\rm pole} e^{i\vartheta^h}\;,
\label{zerlegung}
\ee
can be found in Appendix~C of Ref.~\cite{Ronchen:2014cna} and is identical to the definition given in Ref.~\cite{Workman:2013rca}. The photocoupling at the pole characterizes the coupling of the $\gamma N$ channel to a resonance, independently of the final state in the reaction under consideration. Note that, in general, the complex $\tilde A^{h}_{\rm pole}$ cannot be compared to the real-valued helicity amplitudes $A^{h}$, see Sec.~D of Ref.~\cite{Ronchen:2014cna} for further remarks.

\begin{figure}
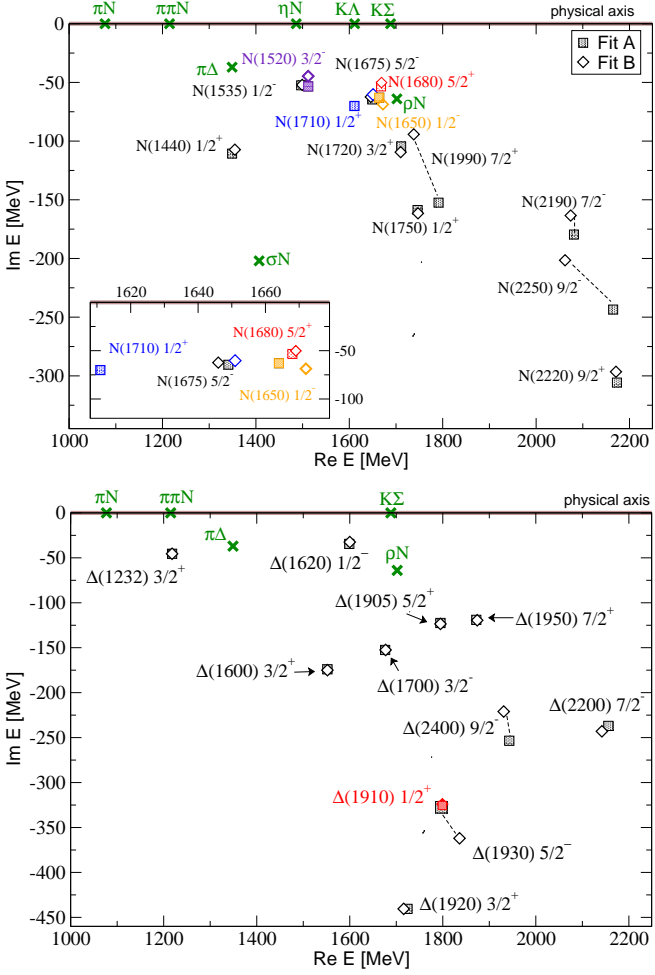
%[htbp]
\begin{center}
\includegraphics[width=1\linewidth]{cmplx_plane_1h_FitA_ct_FitB_ct_f_piNN_Refit.eps} \\ \vspace{0.2cm}
\includegraphics[width=1\linewidth]{cmplx_plane_3h.eps} 
\end{center}
\caption{Pole positions of the isospin $I$=1/2 (above) and $I$=3/2 (below) resonances extracted from fit A (filled squares) and from fit B (empty diamonds). For a better differentiation, the names and pole positions of some resonances are marked with different colors. The green crosses denote the branch points of the amplitude. Note that all cuts, starting at the branch points, are chosen in the negative Im $E$ direction.  }
\label{fig:cmplx_plane}
\end{figure}

In Tables~\ref{tab:poles1} to \ref{tab:photo} we list the pole positions, residues and the photocouplings at the pole of the present study. In
 addition to the values extracted in the current fits A and B we list the pole positions and residues found in fit A of the J\"ulich2012 
 analysis~\cite{Ronchen:2012eg}, called {fit A$_\text{had}$} in the present study, and the photocouplings of fit 2 from the J\"ulich2013 
 analysis~\cite{Ronchen:2014cna}. Note that in the latter study, the parameters of the hadronic $T$-matrix were not altered, i.e. the 
 resonance pole positions and hadronic residues are the same as in fit A$_\text{had}$ of Ref.~\cite{Ronchen:2012eg}. An overview of the pole positions of fit A and B is also given in Fig.~\ref{fig:cmplx_plane}.
 
 In Table~\ref{tab:poles2} the $\pi\Delta$ channel labeled (6) corresponds to the case where $|J-L|=1/2$ and the one labeled (7) to $|J-L|=3/2$. Also the $\rho N$ channel can couple to a resonance with a given $J^P$ in multiple ways, cf. Table~11 in Ref.~\cite{Ronchen:2012eg}. Here, we only quote normalized residues for $\pi\Delta$, since at energies well above the $\pi\Delta$ threshold this channel can be regarded as being composed of the two stable particles $\pi$ and $\Delta$. In general, the resonance coupling at the pole to a channel like $\pi\Delta$ is a function of the center-of-mass
momentum of the stable particle (that equals the summed momenta of the decay products of the unstable particle). Here, we
do not quote this function of $q_{{\rm c.m.}}$ but choose $q_{{\rm c.m.}}$ as on-shell three-momentum of a stable $\Delta$ of
mass $m=1232$~MeV and a pion. Obviously, this prescription does not lead to meaningful results for the very broad $\sigma$ in the $\sigma N
$ channel, or the $\rho N$ channel. In the latter channel, most resonances are not far above the threshold that is situated around
$E=(1.7-i\,0.075)$~GeV, and the $\rho$ cannot be considered a stable particle.
 
   %------------------------------

\subsection{Discussion of specific resonances} 
\label{spec_res}

Compared to the earlier analysis of Ref.~\cite{Ronchen:2012eg}, the extension of the model to eta photoproduction did not require the inclusion of additional bare $s$-channel states,  and we find no new dynamically generated resonances either. In particular, there is no need to include a narrow state at around $E\approx 1.68$~GeV. The narrow structure discovered in eta photoproduction on the neutron~\cite{Kuznetsov:2006kt, Miyahara:2007zz, Werthmuller:2013rba} is absent in the present analysis of eta photoproduction on the proton.

In the following, when discussing selected resonance states, we always refer to the values quoted in Tables~\ref{tab:poles1} to \ref{tab:photo}. 

{$\bf S_{11}$:} While the real part of the pole position of the $\mathbf{N(1535)}$ {\bf 1/2}$^\mathbf{-}$ is very stable throughout all three fits A, B and A$_\text{had}$, the width is by 30~MeV larger in the new fits that include eta photoproduction data. 
The new value is close to the one found in a recent analysis~\cite{Svarc:2014aga} of the GW-SAID WI08 solution~\cite{Workman:2012hx}, where elastic $\pi N$ and $\pi N\to\eta N$ data were fitted. Also the mass in the latter analysis is very similar to our fits.
We obtain the same values for the normalized $\eta N$ residue in all three fits. While the elastic $\pi N$ residue is larger in the new fits, the coupling to this channel is still considerably smaller than the coupling to $\eta N$. In both current fits A and B, the magnitude of the photocoupling $A^h_{\rm pole}$ is more than twice as large as in the previous fit 2 of Ref.~\cite{Ronchen:2014cna} where only pion photoproduction data were considered.
 This change is related to the increase of the width for the $N(1535)$ $1/2^+$, because the resonance width and size of the photocoupling at the pole are strongly correlated. On the whole, comparing the resonance parameters of the $N(1535)$ $1/2^-$ of the older fits to the new ones, the inclusion of eta photoproduction seems to have noticeable impact for this resonance. The influence of the new $T$ and $F$ data~\cite{Akondi:2014ttg}, on the other hand, is rather limited as the parameters in fit A and B do not exhibit major differences. This observation is in agreement with the similarity of fit A and B for $T$ and $F$ at $E\sim 1.5$~GeV (cf. Fig.~\ref{fig:TFetap}~).

By contrast, the pole of the second resonance in the $S_{11}$ partial wave, $\mathbf{N(1650)}$ $\mathbf{1/2^-}$, is located in an energy region where the deficiencies of the prediction of fit A for $T$ and $F$ become more apparent. Accordingly, slightly larger variations are found in the pole positions of fits A and B, and also compared to fit A$_{\rm had}$. Although the width is smaller in the new fits A and B, the resonance is still broader and has a higher mass than in Ref.~\cite{Svarc:2014aga}. In Ref.~\cite{Svarc:2014sqa}, however, the results from a Laurent-Pietarinen (L+P) expansion of the GW-SAID CM12 solution~\cite{Workman:2012jf} for pion photoproduction give a pole position of $E_0=1655(11)-i63.5(8.5)$~MeV, which is closer to our values. 
As in fit A$_\text{had}$, the current fits reveal a strong coupling to the $KY$ channels. The magnitude of the photocoupling to the $N(1650)$ $1/2^-$  is more than twice as large in the fits including eta photoproduction compared to the older fit 2 of Ref.~\cite{Ronchen:2014cna}, where only pion photoproduction was considered. While fit 2 yielded a value smaller than the one of Ref.~\cite{Svarc:2014sqa}, the photocoupling is
now larger in fit A and B. 

In summary, the pole positions of the $N(1535)$ $1/2^-$ and the $N(1650)$ $1/2^-$ come closer to the values of the GW-SAID analysis. As
resonance widths and sizes of helicity couplings are correlated, the values of the latter also approach the values of that analysis. The eta photoproduction data have a strong influence on the $S_{11}$ resonance properties.

{$\bf P_{11}$:} Besides the nucleon pole, we find two resonances in the $P_{11}$ partial wave. One of them, the Roper resonance $N(1440)$ 1/2$^+$ is dynamically generated from the interplay of the $t$- and $u$-channel diagrams. As the fit parameters corresponding to these $T^\npo$ diagrams are not altered in the present study the extracted resonance parameters do not change much. The situation is different for the third state, the $N(1710)$ 1/2$^+$.
%Currently, it is rated with three stars by the particle data group~\cite{Agashe:2014kda}, but in Ref.~\cite{Burkert:2014wea} it was proposed to be promoted to four stars.  
This explicit $s$-channel state was introduced in the J\"ulich model in Ref.~\cite{Ronchen:2012eg} mainly to improve the description of the pion-induced $\eta N$ and $K\Lambda$ channels. Since it couples only weakly to the $\pi N$ channel its resonance parameters are poorly constrained from $\pi N$ elastic scattering. As can be seen in Table~\ref{tab:poles1}, the extension of the fit to new, inelastic reaction channels results in noticeable changes in the pole position. Moreover, also the inclusion of new observables for a specific reaction, here $T$ and $F$ in $\gamma p\to \eta p$ in fit B, leads to significant variations not only in the pole position but also for the residues and photocouplings. The latter observation suggests that additional information from inelastic channels, e.g. in form of new polarization measurements, might help to fix the parameters of the $N(1710)$ 1/2$^+$. 

In all our fits, the $N(1710)$ 1/2$^+$ has a lower mass and is narrower than in recent analyses by the ANL-Osaka ($E_0=1746-i177$~MeV)~\cite{Kamano:2013iva} and the Bonn-Gatchina groups ($E_0=1687\pm17-i(100\pm12.5)$~MeV)~\cite{Anisovich:2011fc}. In the L+P analysis of the GW-SAID CM12 solution in Ref.~\cite{Svarc:2014sqa}, a broad state with a higher mass associated with this resonance  is found that can be alternatively explained as the $\rho N$ complex branch point. However, the authors state that additional information from other decay channels beside $\pi N$ is needed to distinguish between the two options. See also Ref.~\cite{Svarc:2014aga} by the same authors where the same conclusion was drawn. Note that in the present study the $\rho N$ complex branch point is included explicitly. 

In addition to the $N(1440)$ 1/2$^+$ and the $N(1710)$ 1/2$^+$ we find non-conclusive indications for another, very broad and dynamically generated pole at $E\sim1.75$~GeV.

{$\bf P_{13}$:} We include one bare $s$-channel state in the $P_{13}$ partial wave, the $N (1720)$ 3/2$^+$. Although we observed a noticeable sensitivity of the description of the $\eta N$ channel on variations in the $P_{13}$ partial wave, the pole position of the $N (1720)$ 3/2$^+$ is very similar in fits A and B.
The impact of the new $T$ and $F$ data from MAMI can be seen in the photocouplings of this state (fit B vs. fit A). This is reflected in the discrepancies observed in the $M_{1+}$ multipole in Fig.~\ref{fig:mltp_etap}.

In different GW-SAID solutions~\cite{Svarc:2014sqa,Svarc:2014aga} and in the Bonn-Gatchina analysis of Ref.~\cite{Anisovich:2011fc} the $N (1720)$ 3/2$^+$ has a pole position with a real part 20 to 80~MeV lower than in our fits and an imaginary part more than 50~MeV larger. By contrast, the ANL-Osaka group~\cite{Kamano:2013iva} finds values closer to ours.

 We tested the influence of a second explicit resonance state in the $P_{13}$ partial wave but observed no significant improvement of the fit results. 
In this partial wave, the Bonn-Gatchina group finds strong evidence for a state named $N(1900)$ $3/2^+$ in the photoproduction of 
$K\Lambda$ and $K\Sigma$~\cite{Nikonov:2007br,Anisovich:2011fc}. It has also been confirmed in $\gamma p\to K^+\Lambda$ in an 
effective Lagrangian model~\cite{Maxwell:2012zz} and in a covariant isobar-model single channel analysis~\cite{Mart:2012fa}.
It remains to be seen, whether this state is also needed in the J\"ulich approach once the analysis is extended to kaon photoproduction.  Note that the  $N(1900)$ $3/2^+$ is also included in the ANL-Osaka analysis~\cite{Kamano:2013iva} and in the Gie{\ss}en model~\cite{Penner:2002ma}.

{$\bf D_{13}$:} While the real part of the pole positions of the $N (1520)$ 3/2$^-$ is unchanged in fits A and B, the imaginary part is about 10~MeV smaller in fit B. In the previous fit A$_\text{had}$ the real and the imaginary part were similar to fit A, -2Im~$E_0$=110~MeV. In the GWU-SAID solutions analyzed in Refs.~\cite{Svarc:2014sqa,Svarc:2014aga} and the Bonn-Gatchina analysis~\cite{Anisovich:2011fc}, widths of about 110~MeV are also found. With -2Im~$E_0$=78~MeV, a smaller width was extracted in the ANL-Osaka analysis~\cite{Kamano:2013iva}.

Moderate changes in our three fits can also be observed for the values of the residues and photocouplings. Although the
$N (1520)$ 3/2$^-$ is well determined from elastic $\pi N$ scattering and no new information from this channel was included in the new fits, certain changes in the resonance parameters are not surprising. Due to the well-known $SD$-wave interference in the pion-induced $\eta N$ production resulting in a $u$-shape form of the differential cross section (cf. Fig.~\ref{fig:etaN_dsdo}), the $N (1520)$ 3/2$^-$ shows some sensitivity to the parameterization of the $\eta N$ channel. As can be seen in Fig.~\ref{fig:etaN_dsdo} at $E=1509$~MeV and especially at $E=1576$~MeV, the description of the data differs in all three fits.
The energy bin at $E=1576$~MeV is, on the other hand, a prime example for the systematic problems in the data. Data at
backward angles are available with small error bars, but not in agreement with other data spanning the entire angular region.
Underestimated normalization problems can obviously change the angular dependence significantly and have a large impact on the
partial-wave content. Better data are called for.

\begin{figure}
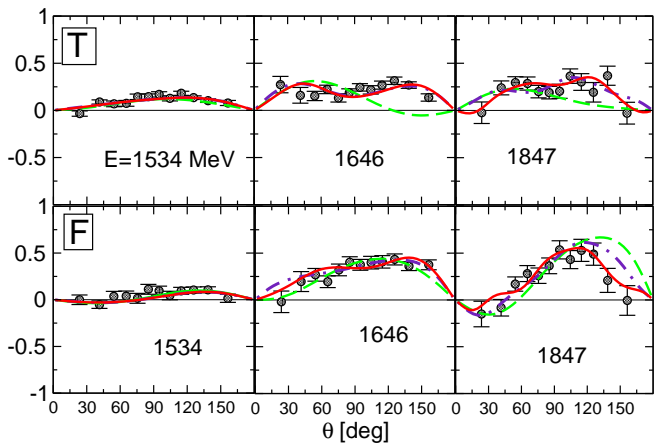
%[htbp]
\begin{center}
\includegraphics[width=1\linewidth]{T_etap_MAMI_A2_PW_2.eps} \\ \vspace{-0.18cm}
\includegraphics[width=1\linewidth]{F_etap_MAMI_A2_PW_2.eps} 
\end{center}
\caption{Partial wave contribution to $T$ (upper row) and $F$ (lower row) in fit B. Solid (red) line: fit B; dashed (green) line: $S_{11}$, $P_{13}$, $D_{13}$ wave only; dash-dotted (indigo) line: all $S$-, $P$-, $D$-, and $F$-waves. }
\label{fig:tf_PW}
\end{figure}

{$\bf D_{15}$, $\bf F_{15}$:} Although the poles of the $N (1675)$ 5/2$^-$ and the $N (1680)$ 5/2$^+$ are located in an energy region where the prediction of fit A for $T$ and $F$ becomes worse (cf. Fig.~\ref{fig:TFetap}), the pole positions and residues exhibit only minor differences in fit A and B. %Regarding the photocouplings, the values extracted in the previous fit 2 deviate stronger from the current fits than fit A and B deviate among themselves.
Still, as can be seen in Fig.~\ref{fig:tf_PW}, in the current fit B, the $D_{15}$ and $F_{15}$ are important to achieve a good description of the new $T$ data in eta photoproduction. Whereas for $F$ a qualitative description of the data is feasible with the $S_{11}$, $P_{13}$, and $D_{13}$ partial waves alone, in case of $T$ all $S$-, $P$-, $D$-, and $F$-waves are needed at medium and higher energies.

In the current form of the approach, only one bare $s$-channel state is incorporated in the $F_{15}$ partial wave.  For a discussion of a possible second explicit resonance we refer the reader to Sec.~4.3 of Ref.~\cite{Ronchen:2012eg}.

{$\bf F_{17}$:} The $\eta N$ normalized residue of the $N (1990)$ 7/2$^+$ is small. Still, the inclusion of the new polarization data for the reaction $\gamma p\to\eta p$ results in a pole position with the real part 53~MeV and the imaginary part almost 60~MeV smaller in fit B than in fit A. The magnitudes of photocouplings, on the other hand, show much less variations in the two new fits compared to the previous fit 2.
As remarked in Sec.~4.2 of Ref.~\cite{Ronchen:2012eg}, from an analysis of elastic $\pi N$ scattering not much evidence can be claimed for this resonance. However, in our current fit B the $F_{17}$ partial wave seems to play a certain role in the parameterization of $T$ in $\gamma p\to\eta p$, cf. Fig.~\ref{fig:tf_PW}. At energies in the range of the pole position of the $N (1990)$ 7/2$^+$, the $F_{15}$ alone plus the $S$-, $P$-, and $D$-waves does not yield a qualitative description of the data. However, evidence for this resonance from the current database is weak in general.

{$\bf G_{17}$, $\bf G_{19}$, $\bf H_{19}$:} We include bare $s$-channel states identified with the $N (2190)$  7/2$^-$, $N (2250)$ 9/2$^-$ and $N (2220)$ 9/2$^+$ resonance. The parameters of those broad states are less stable, as was already observed in the J\"ulich2012 analysis~\cite{Ronchen:2012eg}. 

{\bf Isospin $I=3/2$ resonances}:
Since the $\eta N$ channel does not couple to resonances with $I=3/2$, the pole positions and hadronic residues are very similar in fits A and B. The inclusion of eta photoproduction data can change the $I=3/2$ resonances only indirectly through the mixed-isospin $\pi N$
channels in pion photoproduction, and the mixed-isospin channels in the reaction $\pi N\to K\Sigma$. Nonetheless, the mass of the $\Delta(1930)$ 5/2$^-$ is about 40~MeV higher and the widths about 70~MeV larger in fit B. Smaller differences in the parameters from fits A and B can also be seen in the $\Delta (2200)$ 7/2$^-$ and the
$\Delta (2400)$ 9/2$^-$. The photocouplings at the pole are marginally less stable than the pole positions.

Comparing fit A and fit A$_{\rm had}$ of Ref.~\cite{Ronchen:2012eg}, the influence of pion photoproduction data included in the former fit but not in the latter is visible in the results for some states, as e.g. the $\Delta(1910)$ 1/2$^+$. Note that also the $\Delta(1232)$ 3/2$^+$ changes its pole position slightly.\\

In the analysis of pion photoproduction within the J\"ulich framework~\cite{Ronchen:2014cna} the uncertainties of the extracted photocouplings were estimated from re-fits based on different re-weighted data sets. In Ref.~\cite{Ronchen:2014cna}, all data included in the fit entered with a universal weight of one. In the present study, however, the situation is different. As described in Sec.~\ref{sec:data_fit}, the quality of the hadronic data requires a specific weighting of the various data sets. Moreover, in case of the elastic $\pi N$ channel we fit to the energy-dependent partial-wave amplitudes of the GW-SAID group for which no errors are provided. As a side remark, it should be noted that the error bars of the corresponding single-energy solutions do not provide enough information for correlated $\chi^2$ fits. Furthermore, also for eta photoproduction certain data sets were included with a higher weight. This renders an  error estimation as performed in Ref.~\cite{Ronchen:2014cna} impracticable for the present analysis.

A comprehensive statistical error analysis is complicated by the large number of data points and free parameters, typically inherent in the kind of analysis at hand. Such an analysis is, to our knowledge, not pursued in any of the current DCC approaches and we postpone a rigorous error analysis to future work. Without such an error analysis, the assessment of the significance of certain less well-determined states, like a potential $N(1750)$ $1/2^+$, is not possible.
Concerning the significance of resonance signals, the systematic elimination of states as pursued in Ref.~\cite{Arndt:2006bf} is also a necessary task which we postpone to future work.

%--------------------------

\section{Summary }

Over the last years, measurements of pseudoscalar meson photoproduction reactions with unprecedented quality at facilities like ELSA, MAMI, and JLab have opened a path towards a more complete picture of the baryon spectrum.
The photoproduction of $\eta$ mesons is isospin selective and allows for an analysis of $N^*$ states unaffected by contributions from $\Delta^*$ states. Furthermore, the $\eta N$ final state is physically open for all resonances in the second resonance region and beyond. Eta photoproduction is, thus, a prime reaction for resonance analysis and future complete experiments. Recently, polarization observables with large angular coverage and high statistics have emerged. Among them are the target asymmetry $T$ and the beam-target asymmetry $F$. The latter observable has been measured at MAMI for the very first time in $\gamma p\to\eta p$.

However, even with the measurement of more observables and an improved coverage of the data in angles and energy, a reliable determination of resonance properties requires a combined analysis of reactions with different initial and final states.  
One of these reactions is $\pi^-p\to\eta n$ where, however, the data situation is known to be problematic. To avoid bias, this requires a refit not only of parameters tied to photoproduction, but also of hadronic parameters. Baryon resonance analyses would greatly benefit from a re-measurement of the $\pi^-p\to\eta n$ reaction.

Dynamical coupled-channel (DCC) approaches provide an especially suited tool to combine different reaction channels in a global analysis.
In the present study, we extended the J\"ulich DCC framework to eta photoproduction. 
Based on a simultaneous analysis of nearly 30,000 data points for pion and eta photoproduction off the proton and the world database on the pion-induced reactions $\pi N\to\pi N,\,\eta N,\,K\Lambda,$ and $K\Sigma$, we extracted the spectrum of nucleon and $\Delta$ resonances in terms of pole positions, residues and photocouplings at the pole in an energy regime from $\pi N$ threshold up to $E\sim$2.3~GeV. In the current approach, unitarity and analyticity are respected which is a prerequisite for a reliable determination of the resonance properties. 

Poles and residues were compared to the preceding J\"ulich2012 analysis~\cite{Ronchen:2012eg} in which only hadronic data were considered. The effect of the photoproduction data is most apparent for higher resonances, but also noticeable in case of well established states like the two $S_{11}$ resonances $N(1535)$ $1/2^-$ and $N(1650)$ $1/2^-$ whose widths change when photoproduction data are included. 
Also, some photocouplings at the pole changed in the present analysis compared to the J\"ulich2013 solution~\cite{Ronchen:2014cna} in which only pion photoproduction data, but no eta photoproduction data were considered. 

In order to estimate the influence of the recent MAMI $T$ and $F$ measurements, two different fits were performed, including the new data only in the second fit. Changes in the resonance parameters are predominantly observed for less well established states like the $N(1710)$ $1/2^+$ or higher lying resonances. Smaller but significant changes appear also for well reputed states and particularly for the photocouplings at the pole. 
Moreover, the new data on $T$ and $F$ have a major influence on the multipoles. 

In general, the multipole content of eta
photoproduction is less well established than for pion photoproduction. This calls for further measurements of single- and
double polarization observables.
Upcoming experiments on polarization observables will have significant impact on the resonance spectrum and will help to identify so-called missing states and determine their resonance parameters.

\begin{acknowledgements}
The authors gratefully acknowledge the computing time granted on the supercomputer JUROPA at J\"ulich Supercomputing Centre (JSC). This work is also partially supported by the EU Integrated Infrastructure Initiative
HadronPhysics3 (contract number 283286), by the DFG (Deutsche Forschungsgemeinschaft, 
SFB/TR 16, ``Subnuclear Structure of Matter''), by the DFG and NSFC through the Sino-German CRC
110, by the National Science Foundation, PIF grant No. PHY 1415459, and by the National
Research Foundation of Korea, Grant No.~NRF-2011-220-C00011.
\end{acknowledgements}

\appendix

\section{Renormalization of the nucleon mass and coupling}
\label{renorma3}
The nucleon is included as an $s$-channel state in the $P_{11}$ partial wave. In contrast to the other explicit states in this partial wave the bare mass $m^b_N$ and the bare coupling $f^b_N$ of the nucleon are not free parameters but undergo a renormalization process such that the nucleon pole position and residue to the $\pi N$ channel correspond to the physical values, i.e. $E_0=m_N^{\rm phys}= 938$~MeV and $f_{\pi NN}=0.964$~\cite{Baru:2011bw}. Note that in the present study the nucleon is only allowed to couple to the $\pi N$ channel. Effects of the coupling to other channels with significantly higher threshold energies are small and can be absorbed in the renormalization process.

In Ref.~\cite{Ronchen:2012eg} the renormalization of the nucleon in the presence of two $s$-channel states was illustrated. In the present study, we introduce an additional contact term in the $P_{11}$ partial wave. Hence, the renormalization procedure has to be modified.

For this purpose we define the following reduced self-energies $\tilde\Sigma$ 
\begin{eqnarray}
 \Sigma_{11}&=& (f_N^b)^{2}\Sigma_{11}^{\text{red}}=x^{2}f_{\pi NN}^{2}\,\Sigma_{11}^{\text{red}}=x^2\,\tilde{\Sigma}_{11} \non
 \Sigma_{12}&=& f_N^b\,\Sigma_{12}^{\text{red}}= x \,f_{\pi NN}\,\Sigma_{12}^{\text{red}}=x\,\tilde{\Sigma}_{12}  \non
\Sigma_{21}&=& f_2^b\,\Sigma_{21}^{\text{red}}= x \,f_{\pi NN}\,\Sigma_{21}^{\text{red}}=x\,\tilde{\Sigma}_{21}  \non
\Sigma_{13}&=& f_N^b\,\Sigma_{13}^{\text{red}}= x \,f_{\pi NN}\,\Sigma_{13}^{\text{red}}=x\,\tilde{\Sigma}_{13}  \non
\Sigma_{31}&=& f_3^b\,\Sigma_{31}^{\text{red}}= x \,f_{\pi NN}\,\Sigma_{31}^{\text{red}}=x\,\tilde{\Sigma}_{31}  \ ,
\label{redsig}
\end{eqnarray}
with $x \in \mathbb{R}$ and the bare $\pi NN$ coupling constant $f_N^b$,
\begin{eqnarray}
 f_N^b&=& x\,f_{\pi NN}\; .
\label{fNb1}
\end{eqnarray}
We also define the reduced nucleon resonance vertices $\tilde\Gamma^{a,c}_{\mu;1}$ via
\begin{eqnarray}
\Gamma_{\mu;1}^{a,c}&=&f_{N}^b \Gamma_{\mu;1}^{\text{red}; a,c}=x\,f_{\pi NN}\, \Gamma_{\mu;1}^{\text{red}; a,c}=x\,\tilde{\Gamma}_{\mu;1}^{a,c}\, .
\label{redquan1}
\end{eqnarray}

The nucleon pole position at $E_0=m^{\rm phys}_N$ is given by a zero of the determinant of $D^{-1}$ of Eq.~(\ref{3res}). Using Eqs.~(\ref{redsig}) and (\ref{redquan1}) we obtain an expression for the bare nucleon mass $m_N^b$:
\begin{eqnarray}
m^b_N&=&m^{\rm phys}_N + x^2 \Bigg[-\tilde\Sigma_{11}\label{mbare} \\ \nonumber &&+ \frac{G_3^{-1} \tilde\Sigma_{12}^2 + 
       \tilde\Sigma_{13} (G_2^{-1} \tilde\Sigma_{13} + 
          2 \tilde\Sigma_{12} \Sigma_{23})}{( 
       \Sigma_{23}^2-G_2^{-1} G_3^{-1})}\Bigg]\;.
\end{eqnarray}
Here, we introduced the auxiliary quantities $G_2^{-1}$ and $G_3^{-1}$:
\begin{eqnarray}
G_2^{-1}&=&G_2^{-1}(E)\equiv \left(E - m_2^b-\Sigma_{22}(E)\right)|_{E=m^N_{\rm phys}}\nonumber \\ 
G_3^{-1}&=&G_3^{-1}(E)\equiv m_N-\Sigma_{33}(E)|_{E=m^N_{\rm phys}}\,.
\end{eqnarray}
In Eq.~(\ref{mbare}) all quantities $\Sigma(E)$, $G_2^{-1}(E)$, and $G_3^{-1}(E)$ are evaluated at $E=m^{\rm phys}_N$.

To determine $x$ and thus the bare nucleon coupling $f^b_N$, we exploit that at the nucleon pole the physical residue $(a_{-1})_{\pi N\to \pi N}$ has to agree with the residue of $T^\po$ from Eq.~(\ref{tpo}). The physical residue is given by
\be
(a_{-1})_{\pi N\to \pi N}=\tilde\gamma_{1}^a\,\tilde\gamma_{1}^c\;,
\label{resin}
\ee
where $\tilde\gamma_{1}^{a}$ ($\tilde\gamma_{1}^{c}$) are the bare nucleon vertices calculated at $E=E_0$ with the physical nucleon coupling $f_{\pi NN}$ instead of
the bare coupling (cf. Appendix B.1. of Ref.~\cite{Doring:2010ap}),
\be
\tilde\gamma_{1}^{a}=i\,\sqrt{\frac{3}{8}}\,\frac{f_{\pi NN}\,k}{\pi\, m_\pi}\frac{E_N+\omega_\pi+m_N}{\sqrt{E_N\omega_\pi(E_N+m_N)}} \ ,
\ee
where $k$ is the particle momentum in the center-of-mass frame.
The residue of $T^\po$ of Eq.~(\ref{tpo}) can be calculated as
\be
{\rm Res}_{E_0=m_N^{\rm phys}}T^\po= \frac{{\rm det}D^{-1}}{\partial_E{\rm det}D^{-1}}T^\po\bigg|_{E=m_N^{\rm phys}}
\ee
with $\partial_E:=\frac{\partial}{\partial E}$. We obtain
\begin{widetext}
\begin{eqnarray}
\tilde\gamma_{1}^a\,\tilde\gamma_{1}^c=\frac{{\rm det} D^{-1}}{\partial_E\; {\rm det}\, D^{-1}}
(x\tilde\Gamma^a_{\mu;1},\Gamma^a_{\mu;2},\Gamma^a_{\mu;\text{3}})%\non&&
\;
\left(\begin{matrix}
E-m^b_1-x^2\tilde\Sigma_{11}&&-x\tilde\Sigma_{12}&&-x\tilde\Sigma_{13}\\
-x\tilde\Sigma_{21}     &&E-m^b_2-\Sigma_{22}&&-\Sigma_{23}\\
-x\tilde\Sigma_{3 1}&&-\Sigma_{32} &&m_N-\Sigma_{33}
\end{matrix}
\right)^{-1}%\non&&
\left(
\begin{matrix}
x\tilde\Gamma^c_{\mu;1}\\
\Gamma^c_{\mu;2}\\
\Gamma^c_{\mu;3}
\end{matrix}
\right)\;.
\label{eq:res}
\end{eqnarray}
\end{widetext}
Both sides of Eq.~(\ref{eq:res}) are  evaluated at $E=m_N^{\rm phys}$. Solving Eq.~(\ref{eq:res}) we arrive at an expression for $x$
which depends only on known or fitted quantities. We can calculate the bare mass and coupling of the nucleon by inserting this expression for $x$ in Eqs.~(\ref{fNb1}) and (\ref{mbare}). Setting $\Sigma_{33}=\Sigma_{13}=\Sigma_{23}=0$ we recover the two resonance case, cf. Sec.~2.2 of Ref.~\cite{Ronchen:2012eg}.

The renormalization procedure is performed for each step in the fitting process.

\begin{table*} \caption{Properties of the $I=1/2$ resonances: Pole positions $E_0$ ($\Gamma_{\rm tot}$ defined as -2Im$E_0$), elastic $\pi N$ residues $(|r_{\pi N}|,\theta_{\pi
N\to\pi
N})$, and the normalized residues $(\sqrt{\Gamma_{\pi N}\Gamma_\mu}/\Gamma_{\rm tot},\theta_{\pi N\to \mu})$ of the inelastic reactions $\pi
N\to \mu$ with $\mu=\eta N$, $K\Lambda$, $K\Sigma$.  (*): not identified with PDG
name; (a): dynamically generated. The resonance parameters of fit A$_\text{had}$ were extracted from fit A of Ref.~\cite{Ronchen:2012eg}, where only hadronic data were considered.}
\begin{center}
\renewcommand{\arraystretch}{1.1}
\begin {tabular}{ll|cc|cc|cc|cc|cc} 
%\begin {tabular}{ll| D{.}{}{3} D{.}{}{4}|D{.}{.}{1} D{.}{}{2} |D{.}{.}{1}  D{.}{}{2} |D{.}{.}{1}  D{.}{}{2} |D{.}{.}{1}  D{.}{}{2} |D{.}{.}{1}  D{.}{}{2} } 
\hline\hline
&&\multicolumn{1}{|l}{Re $E_0$ \hspace*{0.8cm} }
& \multicolumn{1}{l|}{-2Im $E_0$\hspace*{0.2cm} }
& \multicolumn{1}{l}{$|r_{\pi N}|$\hspace*{0.2cm}} 
& \multicolumn{1}{l}{$\theta_{\pi N\to\pi N}$ } 
& \multicolumn{1}{|l}{$\displaystyle{\frac{\Gamma^{1/2}_{\pi N}\Gamma^{1/2}_{\eta N}}{\Gamma_{\rm tot}}}$}
& \multicolumn{1}{l|}{$\theta_{\pi N\to\eta N}$\hspace*{0.1cm}}
& \multicolumn{1}{l}{$\displaystyle{\frac{\Gamma^{1/2}_{\pi N}\Gamma^{1/2}_{K\Lambda}}{\Gamma_{\rm tot}}}$} 
& \multicolumn{1}{l|}{$\theta_{\pi N\to K\Lambda}$\hspace*{0.1cm}}
& \multicolumn{1}{l}{$\displaystyle{\frac{\Gamma^{1/2}_{\pi N}\Gamma^{1/2}_{K\Sigma}}{\Gamma_{\rm tot}}}$} 
& \multicolumn{1}{l}{$\theta_{\pi N\to K\Sigma}$}
\bigstrut[t]\\[0.2cm]
&&\multicolumn{1}{|l}{[MeV]} & \multicolumn{1}{l|}{[MeV]} & \multicolumn{1}{l}{[MeV]} & \multicolumn{1}{l}{[deg]} 
& \multicolumn{1}{|l}{[\%]}  & \multicolumn{1}{l|}{[deg]} & \multicolumn{1}{l}{[\%]}  & \multicolumn{1}{l|}{[deg]} &\multicolumn{1}{l}{[\%]} & \multicolumn{1}{l}{[deg]} \\
		  & fit &&&&&&&&
\bigstrut[t]\\
\hline
 $N (1535)$ 1/2$^-$&   	A 	&1497	&105 & 23& $-48$ & 51 & 110 & 5.6 & \hspace{-0.25cm}$-26$ & 8.9 & \hspace{-0.25cm}$-78$	\\
 				&   	B	&1499	& 104 & 22 & $-46$ & 51& 112& 5.0 & 32 & 5.0& \hspace{-0.25cm}$-69$	\\
				& A$_{\text{had}}$ & 1498 & 74 & 17 & $-37$ & 51 & 120 & 7.7 & 68 & 15 &\hspace{-0.25cm}$-74$  \\
				\hline
 	 			
 $N (1650)$ 1/2$^-$&   	A	&1664	& 126 & 31 & $-59$ & 20 & 47 & 21  & \hspace{-0.25cm}$-56$ & 26 & \hspace{-0.25cm}$-76$	\\
 				&   	B	& 1672	& 137 & 37 & $-59$ & 21 & 48 & 20 &\hspace{-0.25cm}$-54$ & 26 & \hspace{-0.25cm}$-74$	\\
				& A$_{\text{had}}$ & 1677 & 146 & 45 & $-43$ & 15 & 57 & 25 & \hspace{-0.25cm}$-46$ & 26 & \hspace{-0.25cm}$-63$  \\
				\hline
 
 $N (1440)$ 1/2$^+_{(a)}$&A	& 1349	&221 & 32 & $-69$ & 5.0 &11 & 4.5 &$\hspace{-0.25cm}-178$  & 1.1 & 140	\\
 				&   	B	& 1355	& 215& 62 & $-98$ & 7.8 & $\hspace{-0.25cm}-27$ & 16 & 145 & 2.7 & 113\\
				& A$_{\text{had}}$ & 1353 & 212 & 59 & $-103$ & 2 & $\hspace{-0.25cm}-40$ & 2 & 156 & 1 &67 \\
				\hline

 $N (1710)$ 1/2$^+$&    	A	&1611	&140  & 2.7 & $-40$ & 6.1 & \hspace{0.cm}175 & 3.7 & \hspace{-0.25cm}$-49$ & 0.9 & \hspace{-0.25cm}$-58$	\\
 				&   	B	&1651	& 121& 3.2 & \hspace{0.25cm}55 & 16& $\hspace{-0.25cm}-180$ & 12 &\hspace{-0.25cm}$-32$ & 0.4 &\hspace{-0.25cm}$-43$\\
				& A$_{\text{had}}$ & 1637 & 97 & 4&$-30$& 24 & \hspace{0.cm}130& 9.4 & \hspace{-0.25cm}$-83$ & 3.9 & \hspace{-0.25cm}$-136$ \\
				\hline
 
 $N (\textit{1750})$  1/2$^+_{(*,a)}$&  A	& 1746  & 317 & 10 & $-133$ & 0.7 &\hspace{-0.25cm}$-85$ & 0.3 & 127 & 1.8 & \hspace{-0.25cm}$-48$\\
 				&   	B	& 1747 & 323 & 14 & $-144$ & 0.2 & 138&0.4& 86& 1.6& \hspace{-0.25cm}$-55$ \\
				& A$_{\text{had}}$ & 1742 & 318 & 8& \hspace{0.25cm}161 & 0.5 & \hspace{-0.25cm}$-140$ & 0.8 &\hspace{-0.25cm} $-170$ & 2.2 & 4 \\
				\hline
 
 $N (1720)$ 3/2$^+$&  	A	& 1711	& 209 & 5.3 &$-45$ & 1.0 & 132 & 0.3 & \hspace{-0.25cm}$-97$ & 2.2 & 96\\
 				&   	B	& 1710	& 219 & 4.2 &$-47$ & 0.7 & 106 & 1.1 &\hspace{-0.25cm}$-70$ & 0.2 & 79 \\
				& A$_{\text{had}}$ & 1717 & 208 & 7 &$-76$ & 1.2 & 98 & 3.1 & \hspace{-0.25cm}$-89$ & 1.7 & 64 \\
				\hline
 
 $N (1520)$ 3/2$^-$&  	A	& 1512	& 107 & 42 &\hspace{-0.cm} $-15$ & 4.6 & 82 & 7.8 & 155 & 8.9 & 140 	\\
 				&   	B	& 1512	& 89 & 37 & \hspace{-0.cm}$-6$ & 2.6 & 95 & 6.9 & 158 & 4.9 & \hspace{-0.25cm}$-41$\\
				& A$_{\text{had}}$ & 1519 & 110 & 42 & \hspace{-0.cm}$-16$ & 3.5 & 87 & 5.8 & 158 & 0.8 & 163\\
				 \hline
 
 $N (1675)$ 5/2$^-$&  	A	& 1649	&  129 & 25 & $-21$ & 6.1 & \hspace{-0.25cm}$-43$ &0.3 & 90 & 2.0 & \hspace{-0.25cm}$-172$\\
 				&   	B	& 1646	& 125  & 24 & $-22$ & 4.4 & \hspace{-0.25cm}$-43$ & 0.1 & 100 & 3.1 & \hspace{-0.25cm}$-175$\\
				& A$_{\text{had}}$ & 1650 & 126 & 24 & $-19$ & 6.0 & \hspace{-0.25cm}$-40$ & 0.3 & \hspace{-0.25cm}$-93$ & 3.3 & \hspace{-0.25cm}$-168$ \\
				\hline
 
 $N (1680)$ 5/2$^+$&   	A	& 1668	&  107 & 37 & $-23$ & 1.0 & 130 & 0.7 & \hspace{-0.25cm}$-96$ & 0.1 & 145\\
 				&   	B	& 1669	& 100 & 34 & $-19$ & 2.7 & 136 & 0.1 & 90 & 0.4 & 148\\
				& A$_{\text{had}}$ & 1666 & 108 & 36 & $-24$ & 0.4 & \hspace{-0.25cm}$-47$ & 0.2 & \hspace{-0.25cm}$-99$ & 0.1 & 141 \\
				\hline

 $N (1990)$ 7/2$^+$&   	A	& 1791	& 305 & 5.9 & $-93$ & 0.6 & \hspace{-0.25cm}$-106$ & 1.6 & \hspace{-0.25cm}$-131$ & 1.2 & 21\\
 				&   	B	& 1738	& 188 & 4.3 & $-70$ & 1.3 & \hspace{-0.25cm}$-82$ & 2.2 & \hspace{-0.25cm}$-111$ & 0.5 & 24\\
				& A$_{\text{had}}$ & 1788 & 282 & 4 & $-84$ & 0.4 & \hspace{-0.25cm}$-99$ & 1.7 & \hspace{-0.25cm}$-123$ & 0.8 & 28 \\
				\hline
 
 $N (2190)$  7/2$^-$&  	A	& 2081	& 359 & 45 & $-27$ & 0.4 & 137 & 2.0 & \hspace{-0.25cm}$-48$ & 1.2 & \hspace{-0.25cm}$-59$\\
 				&   	B	& 2074	& 327 & 35 & $-40$ & 1.6 & 129 & 0.5 & \hspace{-0.25cm}$-51$ & 1.3 & \hspace{-0.25cm}$-69$	\\
				& A$_{\text{had}}$ & 2092 & 363 & 42 & $-31$ & 0.1& \hspace{-0.25cm}$-28$ & 1.9 & \hspace{-0.25cm}$-51$ & 1.3 & \hspace{-0.25cm}$-63$\\
				\hline
 
 $N (2250)$ 9/2$^-$&  	A	& 2165	& 487 & 21 & $-67$ & 0.7 & \hspace{-0.25cm}$-94$ & 0.6 & \hspace{-0.25cm}$-105$ & 0.4 & \hspace{-0.25cm}$-115$\\
 				&   	B	& 2062	& 403 & 8.2 & $-64$ & 1.7 & \hspace{-0.25cm}$-89$ & 0.6 & \hspace{-0.25cm}$-101$ & 0.2 & 70 \\
				& A$_{\text{had}}$ & 2141 & 465 & 17 & $-67$ & 0.6 & \hspace{-0.25cm}$-92$ & 1.1 & \hspace{-0.25cm}$-103$ & 0.3 & \hspace{-0.25cm}$-114$  \\
				\hline
 
 $N (2220)$ 9/2$^+$&    	A	& 2173	& 611& 70 & $-62$ & 0.3 & \hspace{-0.25cm}$-103$ & 1.0 & 60 & 0.1 & \hspace{-0.25cm}$-136$\\
 				&   	B	&2171	&  593 & 62 & $-59$ & 0.4 & \hspace{-0.25cm}$-101$ & 0.7 & 62 & 0.9 & \hspace{-0.25cm}$-128$\\
				& A$_{\text{had}}$ & 2196 & 662 & 87 & $-67$ & 0.1 & 63 & 0.9 & 53 & 0.8 & \hspace{-0.25cm}$-138$\\
\hline\hline
\end {tabular}
\end{center}
\label{tab:poles1}
\end{table*}

\begin{table*}
\caption{Properties of the $I=3/2$ resonances: Pole positions $E_0$ ($\Gamma_{\rm tot}$ defined as -2Im$E_0$), elastic $\pi N$ residues $(|r_{\pi N}|,\theta_{\pi N\to\pi N})$, and
the
normalized residues  $(\sqrt{\Gamma_{\pi N}\Gamma_\mu}/\Gamma_{\rm tot},\theta_{\pi N\to \mu})$ of the inelastic reactions $\pi N\to K\Sigma$
and $\pi N\to\pi\Delta$.  (a): dynamically
generated. The resonance parameters of fit A$_\text{had}$ were extracted from fit A of Ref.~\cite{Ronchen:2012eg}, where only hadronic data were considered.}
\begin{center}
\renewcommand{\arraystretch}{1.1}
\begin {tabular}{ll|cc|cc|cc|cc|cc}  \hline\hline
&&\multicolumn{2}{|l}{Pole position}
 &\multicolumn{2}{|l}{$\pi N$ Residue}
 &\multicolumn{2}{|l}{$K\Sigma$ channel} 
 &\multicolumn{2}{|l|}{$\pi\Delta$, channel (6)}
 &\multicolumn{2}{|l}{$\pi\Delta$, channel (7)}
\bigstrut[t]\\[0.1cm]
&&\multicolumn{1}{|l}{Re $E_0$ \hspace*{0.8cm} }
& \multicolumn{1}{l|}{-2Im $E_0$\hspace*{0.2cm} }
& \multicolumn{1}{l}{$|r_{\pi N}|$\hspace*{0.2cm}} 
& \multicolumn{1}{l}{$\theta_{\pi N\to\pi N}$ } 
%& \multicolumn{1}{|c}{$\Gamma_{\pi N}/\Gamma_{\rm tot}$}\hspace*{0.3cm}
& \multicolumn{1}{|l}{$\displaystyle{\frac{\Gamma^{1/2}_{\pi N}\Gamma^{1/2}_{K\Sigma}}{\Gamma_{\rm tot}}}$}
& \multicolumn{1}{l|}{$\theta_{\pi N\to K\Sigma}$\hspace*{0.1cm}}
& \multicolumn{1}{l}{$\displaystyle{\frac{\Gamma^{1/2}_{\pi N}\Gamma^{1/2}_{\pi\Delta}}{\Gamma_{\rm tot}}}$} 
& \multicolumn{1}{l|}{$\theta_{\pi N\to \pi\Delta}$\hspace*{0.1cm}}
& \multicolumn{1}{l}{$\displaystyle{\frac{\Gamma^{1/2}_{\pi N}\Gamma^{1/2}_{\pi\Delta}}{\Gamma_{\rm tot}}}$} 
& \multicolumn{1}{l}{$\theta_{\pi N\to \pi\Delta}$}
\\
&&\multicolumn{1}{|l}{[MeV]} & \multicolumn{1}{l|}{[MeV]} & \multicolumn{1}{l}{[MeV]} & \multicolumn{1}{l}{[deg]} 
& \multicolumn{1}{|l}{[\%]}  & \multicolumn{1}{l|}{[deg]} & \multicolumn{1}{l}{[\%]}  & \multicolumn{1}{l|}{[deg]} &\multicolumn{1}{l}{[\%]} & \multicolumn{1}{l}{[deg]} \\
		  & fit &&&&&&&&
\bigstrut[t]\\
\hline
 $\Delta(1620)$	1/2$^-$&   	A 	&1599&69 & 17 &$-106$ & 21 & \hspace{-0.25cm}$-106$ & -&- & 57 & 103 	\\
 				&   	B	&1600	& 65	 & 16 & $-104$ & 22 & \hspace{-0.25cm}$-105$ & - &- & 57 & 105\\
				& A$_{\text{had}}$ & 1599 & 71 & 17 & $-107$ & 22 & \hspace{-0.25cm}$-107$ & -&-&57& 102\\
				\hline

 $\Delta(1910)$ 1/2$^+$&   	A 	&1799&651 & 83 &$ -83$ &2.1 & \hspace{-0.25cm}$-129$ & 54 & 130 &-&-	\\
 				&   	B	&1799	& 648 & 90 & $-83$ & 1.9& \hspace{-0.25cm}$-123$ & 58 & 131 & -&-	\\
				& A$_{\text{had}}$ & 1788 & 575 & 56 & $-140$ & 4.7 & \hspace{-0.25cm}$-144$ & 41 & 71 &-&- \\
				\hline

 $\Delta(1232)$ 3/2$^+$	&   	A 	&1218&91& 45 & $-36$ & & & & \\
 				&   	B	&1218	& 92 & 46 & $-36$ & & & &	\\
				& A$_{\text{had}}$ & 1220 & 86 & 44 & $-35$ & & & & \\
				\hline

 $\Delta(1600)$ 3/2$^+_{(a)}$&   	A 	&1552	&	348 & 24 & $-156$ & 14 & \hspace{-0.25cm}$-6.1$ & 33 & 31 & 1.4 & 36 \\
 				&   	B	&1552	& 350	& 23 & $-155$ & 13 & \hspace{-0.25cm}$-5.6$ & 31 & 31 & 1.3 & 29\\
				& A$_{\text{had}}$ & 1553 & 352 & 20 & $-158$ & 11 & \hspace{-0.25cm}$-7$ & 28 & 28 & 1 & 15 \\
				\hline

$\Delta(1920)$	3/2$^+$&   	A 	&1714	&	882 & 36 & \hspace{0.25cm}147 & 15& \hspace{-0.25cm}$-34$ & 6.3 & 132 & 1.3 & \hspace{-0.25cm}$-115$\\
 				&   	B	&1715	&882 	& 38 & \hspace{0.25cm}146 & 17 & \hspace{-0.25cm}$-35$ & 6.9 & 131 & 1.3 & \hspace{-0.25cm}$-115$\\
				& A$_{\text{had}}$ & 1724 & 863 & 36 & \hspace{0.25cm}163 & 16 & \hspace{-0.25cm}$-21$ & 7 & 144 &1 & \hspace{-0.25cm}$-101$ \\
				\hline

 $\Delta(1700)$ 3/2$^-$&   	A 	&1676	&305 & 24 & $-8.4$ & 1.3 & \hspace{-0.25cm}$-148$ & 5.4 & 166 & 39 & 150	\\
 				&   	B	&1677	& 305 & 24 & $-7.3$ & 1.1 &\hspace{-0.25cm}$ -147$ & 5.4 & 166& 39 & 151	\\
				& A$_{\text{had}}$ & 1675 & 303 & 24 & $-9$ & 1.5 & \hspace{-0.25cm}$-150 $& 5 & 166 & 39 & 149 \\
				\hline

 $\Delta(1930)$	5/2$^-$&   	A 	&1797	&655 & 22 & $-155$ & 3.3 & 0.3 & 14 & 30 & 0.6 & 131	\\
 				&   	B	&1836	& 724 & 34 & $-155$ & 4.3 & \hspace{-0.25cm}$-0.5$ & 15 & 30 & 0.9 & 121	\\
				& A$_{\text{had}}$ & 1775  & 646 & 18 & $-159$ & 3.1 & \hspace{-0.25cm}$-3$ & 12 & 26 & $<1$ & -\\
				\hline

 $\Delta(1905)$	5/2$^+$&   	A 	&1795	&245 & 5.3 & $-82$ & 0.2 & \hspace{-0.25cm}$-148$ & 0.8 & 85 & 9.7 & 80	\\
 				&   	B	&1795	& 247	& 5.3 & $-89$ & 0.1 & \hspace{-0.25cm}$-155$ & 0.9 & 64 & 8.7 & 72\\
				& A$_{\text{had}}$ & 1770 & 259 & 17 & $-59$ & 0.5 &\hspace{-0.25cm}$ -142$ & 4 & 130& 34 & 105  \\
				 \hline

 $\Delta(1950)$	7/2$^+$&   	A 	&1872	&238 & 57 & $-32$ & 3.4 & \hspace{-0.25cm}$-87$ & 55 & 131 & 3.7 &\hspace{-0.25cm}$ -91$	\\
 				&   	B	&1874	& 239 & 56 & $-33$ & 3.1 & \hspace{-0.25cm}$-87$ & 54 & 131 & 3.3 &\hspace{-0.25cm}$ -97$	\\
				& A$_{\text{had}}$ & 1884 & 234 & 58 &$ -25$ & 4.0 &\hspace{-0.25cm}$ -78$ & 55 & 139 &3 &\hspace{-0.25cm}$ -84$\\
				\hline

 $\Delta (2200)$ 7/2$^-$&   	A 	&2156	&	474 & 17 & $-48$ & 0.1 & \hspace{-0.25cm}$-92$ & 2.5 & \hspace{-0.25cm}$-137$ & 25 & 115\\
 				&   	B	&2142	&486 & 17 & $-56$ &  0.5 & \hspace{-0.25cm}$-103$ & 2.2 & \hspace{-0.25cm}$-151$ & 23 & 107 	\\
				& A$_{\text{had}}$ & 2147 & 477 & 17 & $-52$ & 0.6 & \hspace{-0.25cm}$-98$ & 2 & \hspace{-0.25cm}$-145$ & 24 & 111 \\
				 \hline

 $\Delta (2400)$ 9/2$^-$&   	A 	& 1943	&	507 & 15 & $-91$ & 1.0 & 28 & 19 & \hspace{-0.25cm}$-105$ & 1.5 & \hspace{-0.25cm}$-33$\\
 				&   	B	&1931	& 442 & 13 & $-96$ & 0.9 & 25 & 18 & \hspace{-0.25cm}$-110$ & 1.2 & \hspace{-0.25cm}$-1.0$	\\
				& A$_{\text{had}}$ & 1969 & 577 & 25 & $-80$ & 1.3 & 40 & 24 & \hspace{-0.25cm}$-98$ & 3 & 1 \\
 	 			
\hline\hline
\end {tabular}
\end{center}
\label{tab:poles2}
\end{table*}

%\begin{minipage}{0.4\textwidth} 

\begin{table*}
\caption{Properties of the $I=1/2$ (left) and $I=3/2$ resonances (right): 
photocouplings at the pole ($A^h_{\rm pole}$, $\vartheta^h$) according to Eq.~(\ref{zerlegung}). (*): not
identified with PDG name; (a): dynamically generated. The photocouplings labeled ``fit 2" were extracted in Ref.~\cite{Ronchen:2014cna}, where parameters of the hadronic $T$ matrix were not altered, i.e. the photocouplings of fit 2 correspond to the pole positions and residues of fit A$_{\rm had}$ in Tables~\ref{tab:poles1} and \ref{tab:poles2}.}
\begin{center}
\renewcommand{\arraystretch}{1.15}
%\begin {tabular}{ll| D{.}{.}{1} D{.}{}{5} | D{.}{.}{1} D{.}{}{5} } 
\resizebox{2.07\columnwidth}{!}{
\begin {tabular}{ll| cc|cc ||  ll |cc|cc} 
\hline\hline
& & \multicolumn{1}{c}{$\mathbf{A^{1/2}_{pole}}$\hspace*{0.2cm}} 
& \multicolumn{1}{c|}{$\mathbf{\vartheta^{1/2}}$ } 
& \multicolumn{1}{c}{$\mathbf{A^{3/2}_{pole}}$\hspace*{0.2cm}} 
& \multicolumn{1}{c||}{$\mathbf{\vartheta^{3/2}}$ } 
& & & \multicolumn{1}{c}{$\mathbf{A^{1/2}_{pole}}$\hspace*{0.2cm}} 
& \multicolumn{1}{c|}{$\mathbf{\vartheta^{1/2}}$ } 
& \multicolumn{1}{c}{$\mathbf{A^{3/2}_{pole}}$\hspace*{0.2cm}} 
& \multicolumn{1}{c}{$\mathbf{\vartheta^{3/2}}$ } 
\bigstrut[t]\\[0.2cm]
&& \multicolumn{1}{c}{{\footnotesize[$10^{-3}$ GeV$^{-\frac{1}{2}}$]}} &\multicolumn{1}{c|}{\footnotesize[deg]} & \multicolumn{1}{c}{\footnotesize[$10^{-3}$ GeV$^{-\frac{1}{2}}$]} & \multicolumn{1}{c||}{\footnotesize[deg]} 
&&& \multicolumn{1}{c}{\footnotesize[$10^{-3}$ GeV$^{-\frac{1}{2}}$]} &\multicolumn{1}{c|}{\footnotesize[deg]} & \multicolumn{1}{c}{\footnotesize[$10^{-3}$ GeV$^{-\frac{1}{2}}$]} & \multicolumn{1}{c}{\footnotesize[deg]} 
 \\
		  & fit 		&&&&& &fit &	&
\bigstrut[t]\\
\hline
 $N (1535)$ 1/2$^-$&   	A & 107  &      4.6   &&& $\Delta(1620)$	1/2$^-$&   	A  & 15 & 13	\\ 	
 				&   	B & 106 & 5.2&&&	&   	B & 14 & 26	\\
				& 	2 & 50$^{+4}_{-4}$ & $-14^{+12}_{-10}$ 	 &&&& 	2 &28$^{+6}_{-2}$ &14$^{+1}_{-4}$ 	\\ \hline
 	 			
 $N (1650)$ 1/2$^-$&   	A	&61   &   \hspace{-0.25cm}$ -19$    &&&	$\Delta(1910)$ 1/2$^+$	&   	A	& 307 & 37	\\
 				&   	B	& 59 & \hspace{-0.25cm}$-14$	&&&&   	B & 321 & 39		\\
				& 	2 	&23$^{+3}_{-8}$ &6.0$^{+28}_{-15}$ &&&	& 	2 & 246$^{+24}_{-47}$ & $-21^{+9}_{-4}$ 	\\ \hline
 
 $N (1440)$ 1/2$^+_{(a)}$&A	& \hspace{-0.25cm}$-49$ &      3.2    &&&  $\Delta(1232)$ 3/2$^+$	&A	& \hspace{-0.25cm}$-118 $& \hspace{-0.25cm}$-7.0$ & \hspace{-0.25cm}$-225$ & 4.0	\\
 				&   	B	& \hspace{-0.25cm}$-60$ & \hspace{-0.25cm}$-23$ &&& 	&   	B	& \hspace{-0.25cm}$-117$ & \hspace{-0.25cm}$-6.6$ & \hspace{-0.25cm}$-226$ & 2.8	\\
				& 	2 	& $-54^{+4}_{-3}$ &5.4$^{+2}_{-5}  $ &&& &	2 &$-114^{+10}_{-3}$& $-8.8^{+4}_{-2}$ & $-229^{+3}_{-4}$ & 3.0$^{+0.3}_{-0.4}$	\\ \hline

 $N (1710)$ 1/2$^+$&    	A	& 7.1 & \hspace{-0.25cm}$-177$	&&& $\Delta(1600)$ 3/2$^+_{(a)}$&    	A	& \hspace{-0.25cm}$-216$ & \hspace{-0.25cm}$-43$ & 345 & \hspace{-0.25cm}$-64$	 \\
 				&   	B	& 20 & \hspace{-0.25cm}$-83$ &&& 			&   	B	&\hspace{-0.25cm}$ -230$ & \hspace{-0.25cm}$-42$ & 332 &\hspace{-0.25cm}$ -71$		\\
				& 	2	& 28$^{+9}_{-2}$ & 103$^{+20}_{-6}$ 	&&&& 	2 	& $-193^{+23}_{-24}$ &$-29^{+9}_{-15}$   & 254$^{+85}_{-86}$ & $-70^{+10}_{-6}$  \\ \hline
				 
 $N (\textit{1750})$  1/2$^+_{(*,a)}$&  A & 7.1 & \hspace{-0.25cm}$-129$ &&& $\Delta(1920)$	3/2$^+$&  A	&\hspace{-0.25cm}$ -188$ & 47 & 508 & 69\\
 				&   	B	&5.0 &\hspace{-0.25cm}$ -36$ &&&	&   	B	 &\hspace{-0.25cm}$ -192$ & 46 & 522 & 67 \\
				& 	2 	& 10$^{+3}_{-6}$ & $-147_{-13}^{+12}$	&&&& 	2 	& $-190^{+50}_{-22}$ & 20$^{+24}_{-11}$   & 398$^{+70}_{-67}$ & 70$^{+4}_{-5}$	  	\\ \hline
								
 $N (1720)$ 3/2$^+$&  	A & 55    &   21    & 38 & 67&  $\Delta(1700)$ 3/2$^-$&  	A	& 128 & 4.8   & 110 & 17 	\\
 				&   	B & 39 & 5.3 & 32 & 66	& &   	B	 & 123 & 1.1  & 124 & 22  	\\
				& 	2 & 51$^{+5}_{-4}$ &57$^{+9}_{-4}$   & 14$^{+9}_{-3}$ &102$^{+29}_{-59}$ &&	2 	& 109$^{+10}_{-10}$ &$-21^{+12}_{-6}$ & 111$^{+27}_{-6}$ & 12$^{+9}_{-11}$		\\ \hline
				 
 $N (1520)$ 3/2$^-$&  	A	& \hspace{-0.25cm}$-43$ & \hspace{-0.25cm}$-5.7$ & 96 & 3.6 &		$\Delta(1930)$	5/2$^-$&  	A	& \hspace{-0.25cm}$-277$ & \hspace{-0.25cm}$-8.1$ & 97 & 27\\
 				&   	B	& \hspace{-0.25cm}$-31$ & \hspace{-0.25cm}$-17$ & 75 & 1.7 &  					&   	B & \hspace{-0.25cm}$-270$ & 33 & 153 & 81	\\
				& 	2	& $-24^{+8}_{-3}$ &$-17^{+16}_{-6}$ &117$^{+6}_{-10}$  & 26$^{+2}_{-2}$	&	& 	2 & $-130^{+73}_{-96}$ &$130^{+77}_{-26}$   & 56$^{+3}_{-151}$ & $-12^{+72}_{-76}$ 		\\ \hline
 
 $N (1675)$ 5/2$^-$&  	A & 21 & 33 & 52 & $-14$	&  $\Delta(1905)$	5/2$^+$&  	A & 41 & 91 & -9.4 & 31\\
 				&   	B & 32 & 36 & 51 & $-9.3$	& 			&   	B & 53 & 89 & -30 &  80	\\
				& 	2 & 22$^{+4}_{-7}$ &49$^{+5}_{-2}$  & 36$^{+4}_{-5}$ &$-30^{+4}_{-4}$  &	& 	2 & 13$^{+13}_{-5}$ &  64$^{+72}_{-36}$& $-72^{+16}_{-16}$ &$-67^{+13}_{-7}$  			\\ \hline
				 
 $N (1680)$ 5/2$^+$&   	A & \hspace{-0.25cm}$-24$ & \hspace{-0.25cm}$-29 $& 105 &\hspace{-0.25cm}$ -11 $& $\Delta(1950)$	7/2$^+$&   	A &  \hspace{-0.25cm}$-66$ & \hspace{-0.25cm}$-23$ & \hspace{-0.25cm}$-86$ & \hspace{-0.25cm}$-20$	\\
 				&   	B & \hspace{-0.25cm}$-22 $& \hspace{-0.25cm}$-28$ & 102 &\hspace{-0.25cm}$ -11$ &				&   	B	&\hspace{-0.25cm}$ -68$ & \hspace{-0.25cm}$-19$ & \hspace{-0.25cm}$-84$ & \hspace{-0.25cm}$-19$ \\
				& 	2 & $-13^{+2}_{-5}$ &$ -42^{+9}_{-18}$ & 126$^{+1}_{-2}$ & $-6.5^{+3}_{-2} $& 	& 	2 	& $-71^{+4}_{-4}$ &$-14^{+2}_{-4}$   & $-89^{+8}_{-7}$ & $-10^{+3}_{-1}$  				\\ \hline
				
 $N (1990)$ 7/2$^+$&   	A & 26 & 8.8 & 30 & 126	& $\Delta (2200)$ 7/2$^-$&   	A & 118 & \hspace{-0.25cm}$-22 $& 171 & \hspace{-0.25cm}$-54 $\\
 				&   	B & 29 & 67 & 33 & 39	&  				&   	B	& 106 & \hspace{-0.25cm}$-23$ & 157 &\hspace{-0.25cm}$ -60 $	\\
				& 	2 & 10$^{+11}_{-6}$ &$-103^{+108}_{-155}$  & 53$^{+23}_{-28}$ & 36$^{+17}_{-4}$ 	& 	& 	2 	& 107$^{+11}_{-20}$ &$-36^{+5}_{-5}$  & 131$^{+24}_{-9}$ &$-67^{+9}_{-5}$  		\\ \hline
				 
 $N (2190)$  7/2$^-$&  	A & \hspace{-0.25cm}$-30$ &\hspace{-0.25cm}$ -21$ & 98 &\hspace{-0.25cm}$ -15$ & $\Delta (2400)$ 9/2$^-$&  	A	& \hspace{-0.25cm}$-40$ & 92 & 79 & \hspace{-0.25cm}$-48$ 	\\
 				&   	B &\hspace{-0.25cm}$ -41$ & \hspace{-0.25cm}$-21$ & 85 &\hspace{-0.25cm}$ -22$	& 			&   	B	& \hspace{-0.25cm}$-34$ & 63 & 54 &\hspace{-0.25cm}$ -75$\\
				& 	2 & $-83^{+7}_{-3}$ &$-11^{+6}_{-2}$  & 95$^{+13}_{-10}$ &$-2.7^{+3}_{-5}$	  && 	2 	&$-128^{+46}_{-12}$ & 118$^{+24}_{-3}$  & 115$^{+42}_{-24}$ & $-40^{+17}_{-28}$		\\ \hline
				 
 $N (2250)$ 9/2$^-$&  	A & 130 & 149 & 89 &\hspace{-0.25cm}$ -2.6$	&\\
 				&   	B & 26 & \hspace{-0.25cm}$-26$ & 119 & \hspace{-0.25cm}$-42 $&	\\
				& 	2 & 90$^{+25}_{-22}$ & 131$_{-11}^{+17}$ & 49$^{+31}_{-19}$ & 171$^{+36}_{-43}$ 	&\\ \hline
				 
 $N (2220)$ 9/2$^+$&    	A	& 141 & 105 & 79 & \hspace{-0.25cm}$-47$ &\\
 				&   	B	& 135 & 114 & 82 & \hspace{-0.25cm}$-41$ &\\
				& 	2 	& 233$^{+84}_{-44}$ & 133$^{+10}_{-6}$   & 162$^{+41}_{-38}$ & $-27^{+26}_{-13}$ &	\\ 
\hline\hline
\end {tabular}
}
\end{center}
\label{tab:photo}
\end{table*}

%\end{minipage}

%================================================================================

\end{document}